\documentclass{jfm}
\usepackage{upmath}
\usepackage{amsmath}
\usepackage{amsfonts}
\usepackage{amssymb}
\usepackage{pgf,pgfarrows}
\usepackage{wasysym}
\usepackage{graphicx}
\usepackage[export]{adjustbox}
\usepackage{subfig}
\usepackage{natbib}
\usepackage{upgreek}
\usepackage{mathrsfs}
\usepackage{float}
\usepackage{caption}
\usepackage{soul}
\usepackage{textcomp}
\usepackage{stmaryrd}
\usepackage{color}
\usepackage{mathtools}
\usepackage{todonotes}

\DeclareMathAlphabet{\mathpzc}{OT1}{pzc}{m}{it}

\ifCUPmtlplainloaded \else
  \checkfont{eurm10}
  \iffontfound
    \IfFileExists{upmath.sty}
      {\typeout{^^JFound AMS Euler Roman fonts on the system,
                   using the 'upmath' package.^^J}
}
      {\typeout{^^JFound AMS Euler Roman fonts on the system, but you
                   dont seem to have the}
       \typeout{'upmath' package installed. JFM.cls can take advantage
                 of these fonts,^^Jif you use 'upmath' package.^^J}
       
      }
  \else
    
  \fi
\fi
\ifCUPmtlplainloaded \else
  \checkfont{msam10}
  \iffontfound
    \IfFileExists{amssymb.sty}
      {\typeout{^^JFound AMS Symbol fonts on the system, using the
                'amssymb' package.^^J}
  
         \let\geq=\geqslant
      }{}
  \fi
\fi
\ifCUPmtlplainloaded \else
  \IfFileExists{amsbsy.sty}
    {\typeout{^^JFound the 'amsbsy' package on the system, using it.^^J}
}
    {\providecommand\boldsymbol[1]{\mbox{\boldmath $##1$}}}
\fi

\def\ee{{\rm e}}
\def\ii{{\rm i}}

\affiliation{
$^1$ Environmental and Geophysical Fluids Group, Department of Mechanical Engineering, Indian Institute of Technology, Kanpur, U.P. 208016, India.}

\pubyear{2012}
\volume{xxx}
\pagerange{xxx--yyy}
\title[Effect of free surface on submerged stratified shear instabilities]{Effect of free surface on submerged stratified shear instabilities}
\author[M.H.~Shete and A.~Guha]
{
M\ls I\ls H\ls I\ls R\ns H\ls.\ns  S\ls H\ls E\ls T\ls E\ls $^{1}$\footnote{Present address: College of Earth, Ocean, and Atmospheric Sciences, Oregon State University, Corvallis, Oregon, USA, 97331.}
\and
A\ls N\ls I\ls R\ls B\ls A\ls N\ns G\ls U\ls H\ls A$^{1}$\footnote{Electronic mail for correspondence: anirbanguha.ubc@gmail.com }
}
\date{?? and in revised form ??}

\begin{document}

\maketitle

\begin{abstract}
 In this paper, we have considered the effects of the shallowness of the domain as well as the air-water free surface on the stratified shear instabilities of the fluid underneath. First, we numerically solve the non-Boussinesq Taylor-Goldstein equation for smooth velocity and density profiles of a model shear layer with a free surface.  
 When the depth of the fluid is relatively shallow compared to the shear layer thickness, the surface gravity waves existing at the free surface come closer to the waves existing in the shear layer. This can lead to resonant wave interactions, making the flow unstable to more varieties of modal instabilities. In order to obtain a deeper understanding of the instability mechanisms, we have performed analytical studies with  broken-line profiles (profiles for which vorticity and density are piecewise constant).
 Furthermore, reduced order broken-line profiles have also been developed, based on which dispersion diagrams  are  constructed. Through these diagrams we have underpinned the resonantly interacting waves leading to each type of instability. 
Two new instabilities have been found; one of them, referred to as the ``surface gravity - interfacial gravity (SG-IG) mode'', arises due to the interaction between a surface gravity wave and an interfacial gravity wave, and would therefore be absent if there is no internal density stratification. The other one - the ``surface gravity - lower vorticity  (SG-LV) mode'', which  arises due to the interaction between a surface gravity wave  and the lower vorticity wave, surpasses  Kelvin-Helmholtz (KH) instability  to become the most unstable mode, provided the system is significantly shallow.  Stability boundary of the SG-LV mode is found to be quite different from that of KH. In fact, KH becomes negligible for relatively shallow flows, while SG-LV's growth rate is significant - comparable to the growth rate of KH for unbounded domains ($\approx 0.18$). Moreover, the SG-LV mode is  found to be analogous to the barotropic mode observed in two-layer quiescent flows. We have found that the effect of a free surface on  Holmboe instability is not appreciable. Holmboe in presence of a free surface is found to be analogous to the baroclinic mode observed in two-layer quiescent flows. Except for Holmboe instability, remarkable differences are observed in all other instabilities occurring in shallow domains when the air-water interface is replaced by a rigid lid. We infer that the rigid-lid approximation is valid  for large vertical domains and should be applied with caution otherwise. Furthermore, we have also shown that if  shear is absent at the free surface, our problem can be modeled using a  Boussinesq type approximation, that is, $\mathcal{O}(1)$ density variations in the inertial terms can still be neglected.

\end{abstract}

\section{Introduction}\label{sec:1}
Flows in the natural environment like lakes, estuaries and oceans are shallow, density stratified and exhibit a free surface. In presence of a background velocity shear, which may arise due to wind forcing, \textcolor{black}{and/or} exchange flow,  it is possible for a stable density stratified flow to become unstable. The resulting instabilities are known as stratified shear instabilities, and a few well known examples are Kelvin-Helmholtz (KH) instability, Holmboe
instability and Taylor-Caulfield (TC) instability. These instabilities are often observed in the pycnocline region (sharp density changes due to salinity, temperature, or both) in natural flows. Instabilities cause  the interfacial gravity waves present at the pycnocline to grow and  break, which  often leads to intense turbulent mixing, and thereby profoundly affect the aquatic environment. 


Shear instabilities arising in the environment have been conventionally modeled in a fluid of infinite vertical extent. The effect of the free surface (interface between air and water) is neglected, hence the Boussinesq approximation can be effectively applied.  Boussinesq approximation neglects the density variation effect in the inertial terms but considers it in the buoyancy term \cite[]{turner1979buoyancy}.
This approximation gives accurate results when the density differences are small compared to the mean background density (for example, slight density differences between warm (fresh) and cold (salty) waters).  Therefore,  Boussinesq approximation may not hold if the effect of the free surface (i.e.\ large density difference between air and water) is taken into account. 

There have been numerous studies on linear stability analyses and direct numerical simulations of Boussinesq stratified shear instabilities, e.g.\ \cite{smyth1988finite}, \cite{smyth1989transition}, \cite{smyth1991instability}, \cite{lawrence1991stability}, \cite{sutherland1992stability}, \cite{alexakis2005holmboe}, \cite{carpenter2007evolution}, \cite{smyth2007mixing}, \cite{carpenter2010identifying}, \cite{guha2013evolution} and \cite{rahmani2014}. These studies have considered \textcolor{black}{domains that are much larger vertically than} the shear layer thickness (so as to emulate an infinite vertical domain).  \textcolor{black}{Hence the free surface  is too far away to play any significant role in the instability processes occurring in the pycnocline. For numerical implementation, a ``rigid lid'' boundary condition at the air-water interface ensures that it has no dynamics.} 
An important step towards understanding flows in natural environment is to take into account \textcolor{black}{the finite vertical extent of the domain}, while still using  Boussinesq approximation.  This is achieved by  considering only the water body (of finite vertical extent) and neglecting the air above. The implementation is similar to that of unbounded (infinite extent) flows mentioned above - rigid lid is used as the upper boundary.  \cite{hazel1972numerical} and \cite{haigh1999symmetric} have shown that the presence of a rigid lid close to the shear layer significantly affects the stability characteristics. 
There have been, however, only a few studies which have considered the effect of the free surface  on submerged shear instabilities \cite[]{longuet1998,bakas2009modal}. Here the fluid below is homogeneous, and similar to the case previously mentioned, the domain extends up to the free surface (air above is neglected). While these studies have definitely made a significant advancement over the rigid lid approximation in capturing the non-trivial effects of the free surface on submerged shear instabilities, the free surface modeling is still an approximate one - the non-Boussinesq effects have not been considered. Since the free surface is indeed non-Boussinesq, involving huge density jump between air and water, it needs to be modeled carefully. The correct equations for a non-Boussinesq interface have been outlined in   \cite{barros2011holmboe} and \cite{heifetz2015stratified}. \cite{barros2011holmboe} have analyzed the non-Boussinesq effects for Holmboe instability while \cite{heifetz2015stratified} have analyzed it for Taylor-Caulfield instability.

\begin{figure}
\centering
\includegraphics[width=28pc]{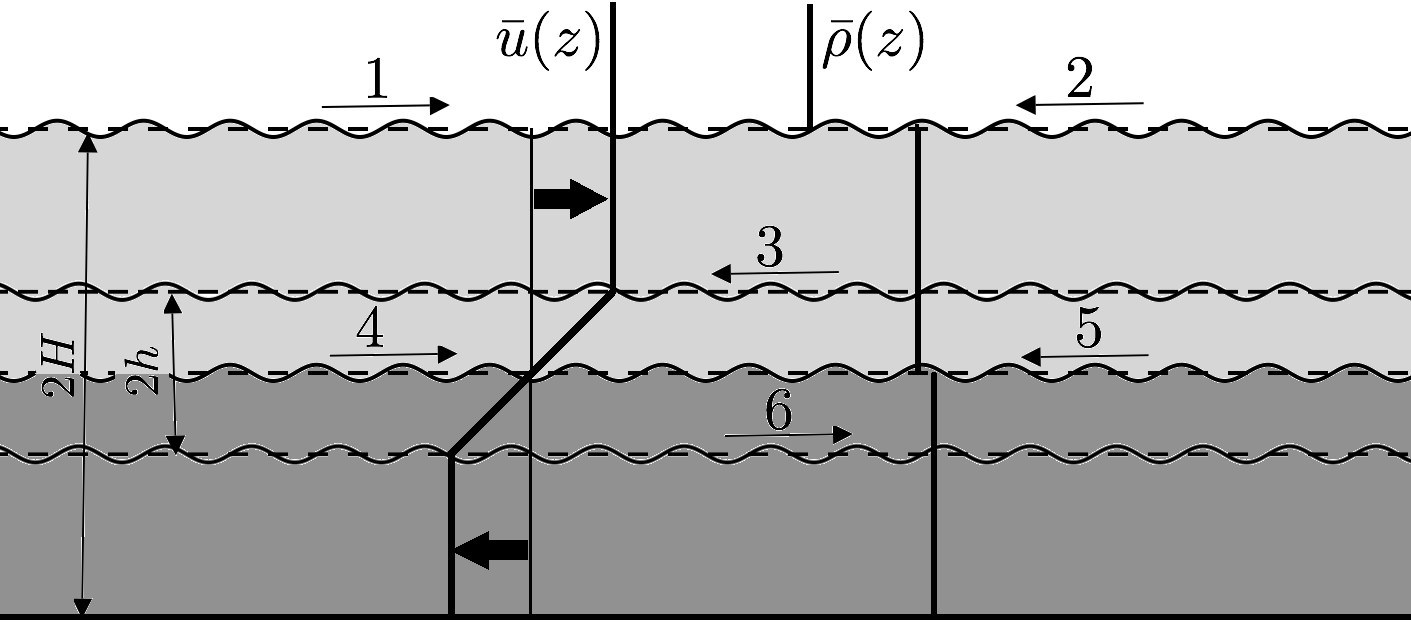}
\caption{
Schematic of a two-layered stratified shear layer in the presence of a free surface. Thick dark arrows indicate the flow direction. Light gray color implies lighter while dark gray color implies heavier fluid. Each thin arrow with a number indicates the  wave present at that location, arrow giving the direction of the intrinsic phase speed.}\label{fig:schematic}
\end{figure}

A key aspect of the free surface is that it supports surface gravity waves. These waves can interact with the different vorticity and interfacial gravity waves that are supported in the stratified shear layer. 
A schematic provided in figure \ref{fig:schematic} shows that there are six waves in the system, two surface gravity waves (marked by $1$ and $2$), two interfacial gravity waves at the pycnocline (marked by $4$ and $5$), and two vorticity waves, one at each vorticity jump (marked by $3$ and $6$).
Mean flow profile of the shear layer  Doppler shifts four out of these six waves (waves $4$ and $5$ are not Doppler shifted since $\bar{u}=0$ there).
Shear instabilities can be conceptually understood in terms of resonant interaction at a distance between  counter-propagating waves \cite[]{holmboe1962behavior,sakai1989rossby,baines1994mechanism,caulfield1994multiple,heifetz1999counter,heif2005,carp2012,guha2014wave}. 
In a counter-propagating system of two waves (each present at its own interface), the intrinsic phase speed of the waves should be opposite to each other. Furthermore, each wave's intrinsic phase speed should be opposite to the local mean flow (unless the local mean velocity is zero). For example,
 classic KH (or Rayleigh) instability will result via an interaction between waves $3$ and $6$ of figure \ref{fig:schematic}, while classic Holmboe will be due to waves $3$ and $4$, as well as $5$ and $6$.
The surface gravity waves can  resonantly interact with (at least) one of the waves existing in the submerged shear layer.  Intuitively, we can expect two additional interactions simply by searching for counter-propagating configurations: (i) waves $2$ and $4$ - surface gravity \textcolor{black}{wave} interacting with interfacial gravity \textcolor{black}{wave} (SG-IG), and (ii) waves $2$ and $6$ - surface gravity \textcolor{black}{wave} interacting with lower vorticity \textcolor{black}{wave} (SG-LV). In this paper we will explore whether these instabilities are actually possible.  
Intuitively, we can also expect the interactions to be more prominent if the free surface is not very far from the pycnocline. We note here in passing that waves $1$ and $3$  give a false impression of  counter-propagating configuration.  Careful observation reveals that the condition for counter-propagation is violated since the phase speed of wave $1$ is \emph{not} opposite to the local mean flow.

The objective of the present study is to consider the effects of shallowness and the free surface on the stratified shear instabilities of the fluid below. The pycnocline is assumed to be sharp, so as to capture both KH and Holmboe modes. In natural settings, the free surface in many occasions is not  far from the pycnocline as compared to the shear layer thickness. Furthermore, natural flows being mostly shallow compared to the shear layer thickness, one can expect a non-trivial effect of the free surface on the instabilities occurring inside the fluid.
For example, while studying  stratified shear instabilities in  Ishikari River estuary
 \cite{yoshida1998mixing} observed KH and Holmboe at the interface (pycnocline) when there is no wind. However in presence of wind the situation is very different -    
``When the wind blows, gravity waves often arise to destroy the interface.  The details of the
complicated mechanisms of this transformation remain undescribed, however, as do those responsible for transport of salt to the surface once the interface is gone''. 



The paper is organized as follows. In \S \ref{sec:2} we outline the fundamental equations describing the flow physics, specifically the non-Boussinesq Taylor-Goldstein equation. We also briefly state the numerical procedure to solve this eigenvalue problem. In \S \ref{sec:3} we perform numerical stability analysis on smooth base state profiles similar to figure \ref{fig:schematic}.
These profiles  provide a good approximation of realistic density stratified shear flows in presence of a free surface. 
The  broken-line profile corresponding to the continuous profile, which is required for providing a mechanistic understanding of the instabilities identified in \S \ref{sec:3}, is discussed in \S \ref{sec:4}.
Reduced order broken-line profiles are devised to single out the key waves that generate different instability mechanisms  due to the presence of the free surface.
Discussions and conclusions are stated in \S \ref{sec:5}.
\section{Governing equations and eigenvalue problem}\label{sec:2}
\subsection{\textcolor{black}{Vorticity equation}}
\label{sec:2.1}
We model the density stratified  shear layer using the $2$D incompressible and inviscid Navier-Stokes equations along with the mass continuity equation. The flow is in the $x-z$ plane with the density stratification along the vertical ($z$) axis.  The horizontal and vertical components of velocity are respectively $u$ and $w$. To accurately capture the effects of the free surface we use a non-Boussinesq model. 
\textcolor{black}{The model takes into account the effect of   density inhomogeneity in the first order of the inertial terms, effects on higher order nonlinear terms are neglected.}
A good way to understand the effect of density inhomogeneity is to examine the vorticity evolution equation, which for a $2$D, inviscid, density stratified fluid  is given by
\begin{equation}\label{eq:2.1}
\frac{\partial q}{\partial t}+u\frac{\partial q}{\partial x}+w\frac{\partial q}{\partial z}=\frac{1}{\rho^{2}}\Big(\frac{\partial \rho}{\partial z}\frac{\partial p}{\partial x}- \frac{\partial \rho}{\partial x}\frac{\partial p}{\partial z} \Big).
\end{equation}
Here vorticity is given by $q \equiv \partial u/\partial z -\partial w/\partial x$, while $\rho$, $p$ and $g$ denote density, pressure and acceleration due to gravity respectively. The terms on the right hand side of (\ref{eq:2.1}) denote the baroclinic generation of vorticity. A base state that varies only along the $z$ axis is assumed, and is  given by $u=\bar{u}(z)$, $w=0$, $\bar{q}(z)=d\bar{u}/dz$, $p=\bar{p}(z)$ and $\rho=\bar{\rho}(z)$. The base state follows hydrostatic pressure balance $d\bar{p}/dz=-\bar{\rho}g$. Perturbations are added to the base flow: $u=\bar{u}(z)+\tilde{u}$, $w=\tilde{w}$,  $p=\bar{p}(z)+\tilde{p}$, $\rho=\bar{\rho}(z)+\tilde{\rho}$, and $q=\bar{q}(z)+\tilde{q}$,  where $\tilde{f}$ denotes the perturbation quantities ($f$ is a placeholder variable). We assume the perturbations to be infinitesimal and linearize (\ref{eq:2.1}). This results in the perturbation vorticity evolution equation 

\begin{equation}\label{eq:2.2}
\frac{\partial\tilde{q}}{\partial t}+\bar{u}\frac{\partial \tilde{q}}{\partial x}=\,\,\,\, \underbrace{-\tilde{w}\frac{d\bar{q}}{dz}}_\text{\clap{Barotropic}}\,\,\,\,\,\,\,\,\,\,\,\,+\,\,\,\,\,\,\,\,\,\,\,\,  \underbrace{\frac{g}{\bar{\rho}}\frac{\partial \tilde{\rho}}{\partial x}}_\text{\clap{Gravitational baroclinic}}\,\,\,\,\,\,\,\,\,\,\,\,\,\,\,\,\,\,+\,\,\,\,\,\,\,\,\,\,\,\,\,\,\,\,\,\, \underbrace{\frac{1}{\bar{\rho}^{2}}\Big(\frac{d \bar{\rho}}{d z}\frac{\partial \tilde{p}}{\partial x} \Big)}_\text{\clap{$T_3\equiv$Non-Boussinesq baroclinic}}.
\end{equation}
Equation (\ref{eq:2.2}) can also be found in \cite{heifetz2015stratified} (their equation (5)). The right hand side of the above equation provides different sources of vorticity generation.
The barotropic generation term arises due to the advection of base state vorticity gradient by the perturbation vertical velocity. 
The \textcolor{black}{``gravitational baroclinic torque''} is responsible for the propagation of interfacial gravity waves at the pycnocline and it is the only baroclinic generation term present when the Boussinesq approximation is invoked.
In the non-Boussinesq (large density stratification) regime, in addition to the \textcolor{black}{gravitational baroclinic generation term}, there is the ``non-Boussinesq baroclinic generation term''(hereafter referred to as $T_3$), that arises out of the density variations in the inertial terms, and is completely independent of the gravitational effects.
However, even for modeling the free surface (the highly non-Boussinesq interface between air and water),  $T_3$ has been neglected in  \cite{bakas2009modal} (see their equation (6)) as well as \cite{longuet1998} (they do not explicitly write in terms of the vorticity equation). In \S \ref{sec:3.2} we provide a detailed discussion so as to delineate the circumstances  under which $T_3$ can or cannot be ignored while modeling interfaces with large density jumps (like the free surface).

\subsection{The non-Boussinesq Taylor-Goldstein equation}
We now present the non-Boussinesq Taylor-Goldstein equation in the inviscid and non-diffusive limit. To obtain it we assume temporal normal mode form for the perturbations given by $\tilde{f}=\hat{f}(z)\ee^{\ii\alpha(x-ct)}$. Here  $\alpha$ and $c$ are respectively the real wavenumber and the complex phase speed ($c = c_{r}+\ii c_{i}$), and $f$ could represent  $u$, $w$, $p$, $q$ or $\rho$. Temporal normal mode form is substituted in the linearized Navier-Stokes equations, yielding 
\begin{equation}\label{eq:2.3}
\underbrace{\bar{\rho}^{\prime}\big[(\bar{u}-c)\hat{w}^{\prime}- \bar{u}^{\prime}\hat{w}\big]}_\text{\clap{Non-Boussinesq}} - \underbrace{\frac{g\bar{\rho}^{\prime}}{\bar{u}-c}\hat{w}}_\text{\clap{Gravitational}} \,\,\, + \,\,\, \bar{\rho}\big[(\bar{u}-c)(\hat{w}^{\prime\prime}-\alpha^{2}\hat{w})\big] - \underbrace{\bar{\rho}\bar{u}^{\prime\prime}\hat{w}}_\text{\clap{Barotropic}} =0.
\end{equation}
Total derivative with respect to $z$ is denoted by $^{\prime}$.  The non-Boussinesq Taylor-Goldstein equation obtained \textcolor{black}{is same as the one obtained by \cite{barros2011holmboe}, \cite{barros2014elementary}, and \cite{carpenter2017}}. 
Terms denoted by the braces indicate the corresponding terms in the perturbation vorticity equation (\ref{eq:2.2}).
We intend to solve (\ref{eq:2.3}) for smooth profiles of base state velocity and density, and for this numerical eigenvalue solver is necessary. The numerical solution of (\ref{eq:2.3}) poses a resolution issue. A very fine spacing in parameter space is required which raises the computational cost tremendously. An efficient way to control the resolution issue is to introduce numerical viscosity ($\mu$) and diffusivity ($\kappa$). Therefore we numerically solve the viscous diffusive form of the non-Boussinesq Taylor-Goldstein equation, which is given by
\begin{subequations}
\begin{equation}
 \bar{\rho}^{\prime}[(\bar{u}-c)\hat{w}^{\prime}-\bar{u}^{\prime}\hat{w}]+\bar{\rho}[(\bar{u}-c)(\hat{w}^{\prime\prime}-\alpha^{2}\hat{w})- \bar{u}^{\prime\prime}\hat{w}]=-\ii\alpha\hat{\rho}g-\frac{\ii}{\alpha}\mu[\hat{w}^{\prime\prime\prime\prime}-2\alpha^{2}\hat{w}^{\prime\prime}+\alpha^{4}\hat{w}], \label{eq:2.4} 
\end{equation}
\begin{equation}
 \ii \alpha(\bar{u}-c) \hat{\rho}+\hat{w}\bar{\rho}^{\prime}=\kappa(\hat{\rho}^{\prime\prime}-\alpha^{2}\hat{\rho}).\label{eq:2.5}
\end{equation}
\end{subequations}
A complete derivation of (\ref{eq:2.4})-(\ref{eq:2.5}) is provided in Appendix \ref{appA}. 
The Boussinesq limit of the above equation set can be found in the appendix of \cite{smyth2011narrowband}.
In the next subsection we briefly describe the numerical strategy to solve the eigenvalue problem (\ref{eq:2.4})-(\ref{eq:2.5}).
\subsection{Solution of the eigenvalue problem}
Equation set (\ref{eq:2.4})-(\ref{eq:2.5}) can be converted into  a generalized eigenvalue problem of the form $M \boldsymbol{\vartheta}=c N\boldsymbol{\vartheta}$ :

\begin{equation}\label{eq:2.6}
\begin{bmatrix} 
M_{11} & M_{12} \\
M_{21} & M_{22} 
\end{bmatrix}
\begin{bmatrix}
\vartheta_{1} \\
\vartheta_{2}
\end{bmatrix}
=c
\begin{bmatrix}
N_{11} & N_{12} \\
N_{21} & N_{22} 
\end{bmatrix}
\begin{bmatrix}
\vartheta_{1} \\
\vartheta_{2}
\end{bmatrix}.
\end{equation}
Elements of the matrices are given by 
\begin{equation*}
M_{11}=\bar{\rho}^{\prime}[-\bar{u}D + \bar{u}^{\prime}] + \bar{\rho}[-\bar{u}D^{2} + \alpha^{2}\bar{u} + \bar{u}^{\prime\prime}]- \frac{\ii}{\alpha}\mu[D^{4}-2\alpha^{2}D^{2} + \alpha^{4}],\,\,\,\,\,\,\,
\end{equation*}
\begin{equation*}
M_{12}=-\ii\alpha g,\,\,\,\,M_{21}=\bar{\rho}^{\prime},\,\,\,\,M_{22}=\ii\alpha \bar{u} - \kappa[D^{2}-\alpha^{2}],\,\,\,\,\vartheta_{1}=\hat{w},\,\,\,\,\vartheta_{2}=\hat{\rho},\,\,\,
\end{equation*}
\begin{equation*}
N_{11}=-\bar{\rho}^{\prime}D+\bar{\rho}[-D^{2}+\alpha^{2}],\,\,\,\,N_{12}=0,\,\,\,\,N_{21}=0,\,\,\,\,N_{22}=\ii\alpha.\,\,\,\,\,\,\,\,\,\,\,\,\,\,\,\,\,\,\,\,\,\,\,\,\,\,\,
\end{equation*}
Equation (\ref{eq:2.6}) is an eigenvalue problem for eigenvalue $c$ and eigenfunctions $\hat{w}$ and $\hat{\rho}$.
The first, second and fourth derivative matrices of the total derivatives with respect to $z$ are given by $D$, $D^{2}$ and $D^{4}$ respectively, which have been numerically evaluated using the fourth order central difference scheme. Since central differencing is not possible at the boundaries, the boundary points have been discretized using second order one sided finite difference scheme.
The boundary conditions used for $\hat{w}$ are impenetrability and free-slip. The impenetrable boundary condition arises due to the continuum hypothesis, and is given by 
\begin{equation}\label{eq:2.7}
\hat{w}=0.
\end{equation}
Furthermore, the free-slip boundary condition,
\begin{equation*}
\frac{d\hat{u}}{dz}=0,
\end{equation*}
along with the fluid being incompressibility gives,
\begin{equation}\label{eq:2.8}
\frac{d^{2}\hat{w}}{dz^{2}}=0.
\end{equation}
We use  insulating boundary condition  for $\hat{\rho}$, which is given by
\begin{equation}\label{eq:2.9}
\frac{d\hat{\rho}}{dz}=0.
\end{equation}
The insulating boundary condition is preferable in cases where there is no physical boundary, since it has minimal effect on the flow.
\textcolor{black}{ For the upper boundary, we have assumed an ``imaginary rigid lid'' in the air region at a certain height (chosen such that it does not adversely impact the results) above the free surface. Boundary conditions (\ref{eq:2.7})-(\ref{eq:2.9}) have been applied to that imaginary rigid lid.}

The matrix eigenvalue problem is solved using  built-in functions in MATLAB. The numerical solution procedure used is similar to the one used by \cite{smyth2011narrowband}. The numerical routine developed has been validated against the Boussinesq shear layer problem of \cite{smyth1988finite} and that of surface gravity waves.
Our primary interest is to capture inviscid instabilities like KH and Holmboe, hence viscosity ($\mu$) and diffusivity ($\kappa$) introduced in (\ref{eq:2.4})-(\ref{eq:2.5}) are treated as purely numerical parameters for controlling the resolution issue. The stability characteristics do not vary appreciably on changing the numerical parameters by an order of magnitude.
 
 \begin{figure}
\centering
\includegraphics[width=32pc]{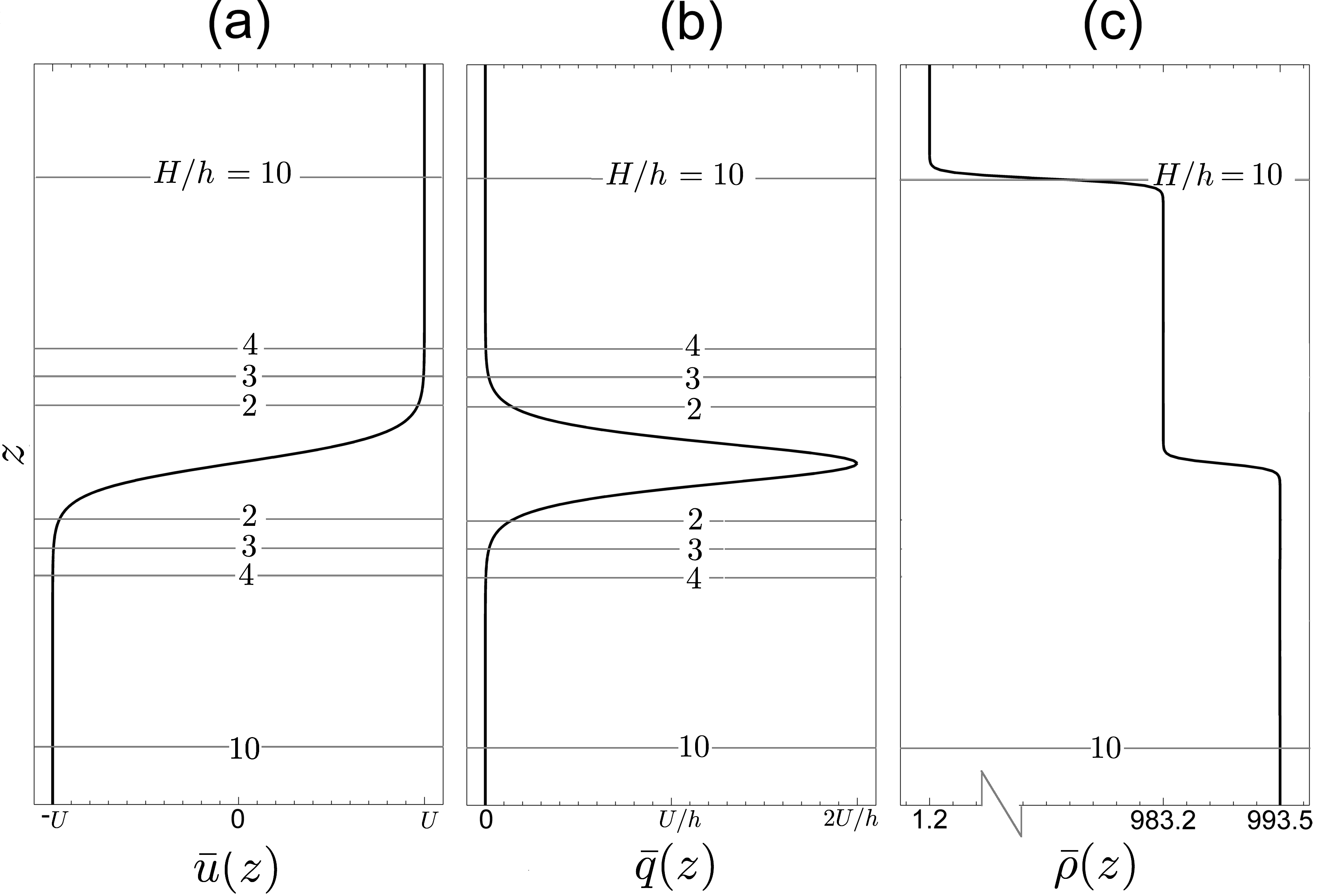}
\caption{The base state plots for (a) horizontal velocity, $\bar{u}$ in m/s, (b) corresponding vorticity, $\bar{q}(z)=d\bar{u}/dz$ in s$^{-1}$, and (c) density, $\bar{\rho}$ in kg/m$^3$ corresponding to $H/h=10$ (which indicates the location of the free surface).  The horizontal lines indicate the domain extent of the water body for each $H/h$. For example, when $H/h=3$, the vertical domain of the water body is confined between the horizontal lines indicated by ``$3$'' (which means that the free surface and the bottom boundary will respectively occur at the upper and lower horizontal lines indicated by ``$3$''). }\label{fig:base_state_plot}
\end{figure}

%
%
%


\section{Numerical stability analysis of the smooth profiles}\label{sec:3}
\subsection{Base state profiles}
In order to model a stratified shear layer, we use hyperbolic tangent functions in $z$ to represent the base state velocity and density profiles.
 Since we are interested in analyzing the effect of the free surface on the shear layer below, we need to extend the velocity profile in the air region. This extension has been made such that (i) the velocity profile is continuously differentiable at the free surface, and (ii) the velocity remains nearly constant with $z$ away from the free surface. 
The first point ensures that the derivatives of base state velocity remain finite, which is required for the numerical stability of (\ref{eq:2.6}). The second point ensures avoiding the  formation of critical layers in the air region. In other words, we avoid the wind-wave instability of \cite{miles1957generation}. The dimensional base state velocity variation is given by:
\textcolor{black}{
\begin{equation}\label{eq:3.1.1}
\bar{u}(z)=U\tanh\bigg(\frac{z-\frac{1}{2}H}{h}\bigg).
\end{equation}
}
Here $U$ (in m/s) represents the surface current and $h$ is the half shear layer thickness. $H$ is the dimensional half-depth of the channel (i.e.\, distance between free surface and bottom is $2H$). In this paper we have taken $h=1/2$ m. 
The base state velocity is depicted in figure \ref{fig:base_state_plot}(a), while the base state vorticity is depicted in figure \ref{fig:base_state_plot}(b).
If $H/h$ is not large, moderately long waves inside the shear layer can feel the effect of the free surface.  
In the present study we have detailed the results for $H/h=10,\,4,\,3$ and $2$. These values of $H/h$ have been chosen so that a clear transition in stability characteristics can be seen as the surface gravity waves come closer to the other waves present in the shear layer (therefore getting an opportunity to resonantly interact with them).
The dimensional base state density profile is given by
\textcolor{black}{
\begin{equation}\label{eq:3.1.3}
\bar{\rho}(z) = \left\{
        \begin{array}{cc}
           \rho_{01}\Bigg(1 - A_{t,\,0}\tanh\bigg[5\bigg(\dfrac{z-H}{h}\bigg)\bigg] \Bigg) & \quad 1.2H\geq z \geq 0.75H, \\ \\
            \rho_{12}\Bigg(1 - A_{t}\tanh\bigg[5\bigg(\dfrac{z-\frac{1}{2}H}{h}\bigg)\bigg] \Bigg)  & \quad 0.75H \geq z\geq 0.
        \end{array}
    \right.
\end{equation}
Here $\rho_{01}=(\rho_{ww}+\rho_{air})/2=491 \,\, \textrm{kg}/\textrm{m}^{3}$ is the mean density of air and warm water, while the Atwood number corresponding to the air - warm water interface (i.e.\ free surface) is given by $A_{t,\,0}=(\rho_{ww}-\rho_{air})/(\rho_{ww}+\rho_{air})=9.97 \times 10^{-1} \approx 1$. Similarly the mean density for cold water and warm water is given by $\rho_{12}=(\rho_{cw}+\rho_{ww})/2=988.35 \,\, \textrm{kg}/\textrm{m}^{3}$ and the Atwood number corresponding to the pycnocline is given by 
\begin{equation}\label{eq:3.2.2}
A_t=\frac{\rho_{cw}-\rho_{ww}}{\rho_{cw}+\rho_{ww}}=\frac{10.3 \, \textrm{kg}\textrm{m}^{-3} }{(993.5+983.2)\,\textrm{kg}\textrm{m}^{-3}} \approx 0.005.
\end{equation}
The densities of cold water and warm water are respectively given by $\rho_{cw}$ and $\rho_{ww}$.
}
Base state density has the units of $\mathrm{kg}$/$\mathrm{m^{3}}$, and is plotted in figure \ref{fig:base_state_plot}(c).

A key parameter of importance is the bulk Richardson number, $J$, given by
\textcolor{black}{
\begin{equation}\label{eq:3.2.1}
J=2A_t\frac{gh}{U^{2}}.
\end{equation}
}
For reporting stability characteristics we have non-dimensionalized the wavenumber, $\alpha$, the phase speed, $c$, and the growth rate $\Im\{\alpha c\}$. The wavenumber is non-dimensionalized by shear layer half-thickness, $h$, while the phase speed is non-dimensionalized by the surface velocity, $U$. Growth rates are non-dimensionalized by the shear scale $U/h$.


\subsection{ Is ignoring the non-Boussinesq baroclinic term legitimate for modeling the air-water free surface?}
\label{sec:3.2}
As noted in \S \ref{sec:2.1}, a couple of previous studies \cite[]{longuet1998,bakas2009modal} that considered the effect of the free surface on the shear instability underneath have ignored the term $T_{3} \equiv (\bar{\rho})^{-2}(d \bar{\rho}/d z) (\partial\tilde{p}/\partial x)$ appearing in (\ref{eq:2.2}). 
 For the purpose of comparison we have analyzed the free surface effect using two different models,  namely the ``Boussinesq free surface model''  (which ignores $T_{3}$) and the ``non-Boussinesq free surface model'' (which fully solves (\ref{eq:2.2})). Here we try to understand the implications of assuming $T_3=0$. There are two possible interpretations of $T_3=0$: 
 
 (i)  $d\bar{\rho}/d z=0$: This is a valid approximation in the Boussinesq limit (where $\bar{\rho} \approx \textrm{constant}$). In our setting, $T_3$ will only matter at the free surface, where obviously $d\bar{\rho}/d z \neq 0$, rather is approximately a delta function. Hence assuming $d\bar{\rho}/d z=0$ for the air-water density jump is  not correct. 
 
 (ii) $\partial\tilde{p}/\partial x=0$: In the inviscid limit
 \begin{equation}
 \frac{\partial \tilde{p}}{\partial x} =-\bar{\rho}\left[\frac{\mathbb{D}\tilde{u}}{\mathbb{D} t} + \tilde{w}\bar{q}\right],
  \label{eq:11111}
 \end{equation}
 where $\bar{q}=d\bar{u}/dz$ and $\mathbb{D}\tilde{u}/\mathbb{D}t \equiv \partial \tilde{u}/\partial t + \bar{u}\partial \tilde{u}/\partial x$ is the linearized material derivative of $\tilde{u}$; see (\ref{eq:A12}). For surface gravity waves, the free surface is a vortex sheet ($\tilde{u}$ changes sign above and below the free surface). To elaborate this point, consider a two-fluid system, upper fluid with  density $\rho_1$ and  lower fluid with density $\rho_2$. For analytical simplicity, both fluids are assumed to be infinitely deep. In the absence of any background flow, the velocity just below (above) the density interface is 
 \begin{equation*}
  \tilde{u}_{\pm}=\pm \omega \mathrm{e}^{\ii (\alpha x- \omega t)},
 \end{equation*}
 where $+(-)$ indicates  below (above), and $\omega=\sqrt{g \alpha(\rho_2-\rho_1)/(\rho_1+\rho_2)}$ is the frequency; see \citet[Chapter 7.7]{kundu2012chapter}. The  velocity at the interface is the average of the two: $\tilde{u}=(\tilde{u}_{+}+\tilde{u}_{-})/2=0$.
Substitution of $\tilde{u}=0$ in  (\ref{eq:11111}) yields  $\partial\tilde{p}/\partial x=0$, \emph{provided} $\bar{q}=0$. Hence $T_3=0$. Proceeding further and using the normal mode ansatz, we obtain the  dispersion relation for surface gravity waves. We note here in passing that such intricacies do not appear in the standard technique for deriving the surface gravity wave dispersion relation, which simplifies the problem from the beginning by using potential flow approximation in each layer, and then makes use of   kinematic and dynamic boundary conditions; see \citet[Chapter 7.2]{kundu2012chapter}. 
However in a generalized scenario where background shear is present at the density interface (i.e.\ $\bar{q}\neq 0$), (\ref{eq:11111}) directly reveals that $\partial\tilde{p}/\partial x =- \bar{\rho} \bar{q} \tilde{w}\neq 0$, implying $T_3 \neq0$. 
In summary, $T_{3}=0$, as implicitly demanded by the Boussinesq free surface model, is not correct when background shear is present at the free surface. 
\subsection{Variation of stability characteristics with $H/h$}

\begin{figure}
\centering{\includegraphics[width=5.4in]{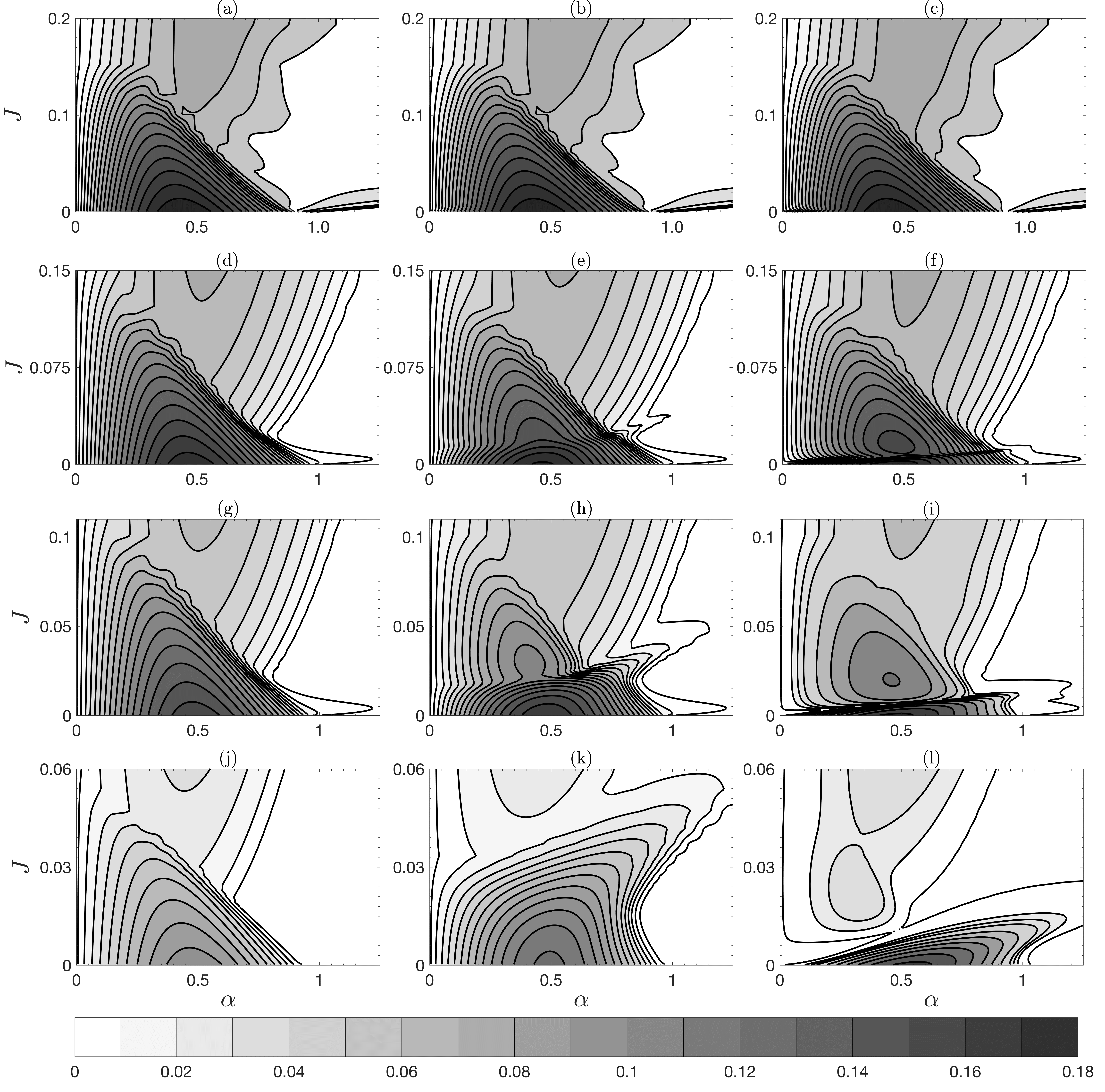}}
  \caption{Growth rate contours in the $\alpha$-$J$ plane for the numerical stability of smooth profiles. Each row represents growth rates for a given $H/h$. First row: $H/h=10$, second row: $H/h=4$, third row: $H/h=3$, and fourth row: $H/h=2$. The left, middle and right columns respectively correspond to rigid-lid, Boussinesq free surface, and  non-Boussinesq free surface.Thus it}
\label{fig:3.2.2}
\end{figure}

Here we have studied the effect of varying the distance between the free surface and the pycnocline (i.e.\ $H/h$)  on the submerged stratified shear layer. As mentioned previously, the proximity between the surface gravity waves, and the vorticity and interfacial gravity waves existing at the shear layer can  affect the stability characteristics.
A comparison of stability characteristics of a stratified shear layer in presence of a rigid lid, a Boussinesq free surface, and an actual/non-Boussinesq free surface for $H/h=10$ are  shown in figures \ref{fig:3.2.2}(a)-(c). 
In this case we see a very good match between the rigid lid, \textcolor{black}{the Boussinesq free surface} and the non-Boussinesq free surface, which is expected since the free surface is located quite far. The maximum growth rate and the corresponding phase speed also agree very well, as can be seen from table \ref{tab:first}.

If we look at the first column of figure \ref{fig:3.2.2}, which corresponds to the rigid-lid case, it becomes clear that decreasing $H/h$ significantly decreases the maximum growth rate. The KH instabilities (the downward pointing closed curves having higher growth rates) are more significantly affected than the Holmboe modes (the upward pointing open curves with lesser growth rates).  Moreover, short waves are stabilized by this process, as is evident from the leftward shift of the right stability boundary. The results are in good agreement with that  of \cite{hazel1972numerical} and \cite{haigh1999symmetric}.


The results obtained on replacing the rigid lid by a non-Boussinesq free surface are significantly different; see the last column of figure \ref{fig:3.2.2}. Decreasing $H/h$ has little effect on the maximum growth rate; see table \ref{tab:first}. However, the stability boundaries change quite dramatically, and furthermore, new modes appear. The distinct identity of the KH mode observed for $H/h=10$ is lost in $H/h=3$; a part of it separates out as ``bubble'' and becomes a part of the Holmboe branch. The other part near $J=0$ axis leads to a new mode, the ``SG-LV mode'', which is distinctly visible for $H/h=3$ and $2$. The SG-LV mode always has the highest growth rate. Comparing with the rigid lid case (first column), it can be straightforwardly argued that the SG-LV mode wouldn't exist if there were no surface gravity waves. We will show later that the SG-LV mode is a result of the interaction between waves $2$ and $6$ in figure \ref{fig:schematic}.

\begin{table}
  \begin{center}
\def~{\hphantom{0}}
  \begin{tabular}{lcccccc}
      Case    & $H/h$ & Mode & $\alpha$ & $J$ & $\gamma$ & $c_{r}$    \\[3pt]
       Rigid lid   & $10$ & KH & $0.4474$ & $0$ & $0.1893$ & $0$    \\
       \,  & $10$ & H & $0.6773$ & $0.3498$ & $0.0781$ & $\pm 0.4140$ \\
       \,  & $4$ & KH & $0.4724$ & $0$ & $0.1769$ & $0$ \\
       \,  & $4$ & H & $0.6815$ & $0.3498$ & $0.0768$ & $\pm 0.4155$ \\
       \,  & $3$ & KH & $0.4892$ & $0$ & $0.1566$ & $0$ \\
       \,  & $3$ & H & $0.6898$ & $0.3498$ & $0.0737$ & $\pm 0.4191$ \\
       \,  & $2$ & KH & $0.4849$ & $0$ & $0.0966$ & $0$ \\
       \,  & $2$ & H & $0.8152$ & $0.5106$ & $0.0618$ & $\pm 0.4864$ \\
     Boussinesq free surface & $10$ & KH & $0.4431$ & $0$ & $0.1893$ & $-0.0001$    \\
       \,  & $10$ & H & $0.6775$ & $0.0.3498$ & $0.078$ & $-0.4146$ \\
       \,  & $4$ & SG-LV/KH & $0.4641$ & $0$ & $0.1816$ & $-0.0153$ \\
        
         \,  & $4$ & H & $0.6856$ & $0.3498$ & $0.0767$ & $-0.4158$ \\
          \,  & $3$ & SG-LV & $0.4850$ & $0$ & $0.1677$ & $-0.0270$ \\
           \,  & $3$ & KH & $0.4139$ & $0.03004$ & $0.1035$ & $-0.0034$ \\
            \,  & $3$ & H & $0.6982$ & $0.3498$ & $0.0734$ & $-0.4198$ \\
             \,  & $2$ & SG-LV & $0.4933$ & $0$ & $0.1224$ & $-0.0375$ \\
            
            \,  & $2$ & H & $0.8738$ & $0.5106$ & $0.05931$ & $-0.4897$ \\
    Non-Boussinesq free surface & $10$ & KH & $0.4431$ & $0$ & $0.1894$ & $-0.0051$    \\
  \,  & $10$ & H & $0.6775$ & $0.3498$ & $0.0788$ & $-0.4155$ \\  
  \,  & $4$ & SG-LV & $0.4557$ & $0$ & $0.1884$ & $-0.0513$ \\
  \,  & $4$ & KH & $0.46826$ & $0.0152$ & $0.15445$ & $-0.00695$ \\
  \,  & $4$ & H & $0.6898$ & $0.3498$ & $0.07708$ & $-0.41638$ \\
   \,  & $3$ & SG-LV & $0.4808$ & $0$ & $0.1844$ & $-0.1010$ \\  
   \,  & $3$ & KH & $0.4599$ & $0.019$ & $0.1209$ & $-0.0073$ \\
   \,  & $3$ & H & $0.715$ & $0.3498$ & $0.0729$ & $-0.4197$ \\
  \,  & $2$ & SG-LV & $0.5477$ & $0$ & $0.1658$ & $-0.1789$ \\
  \,  & $2$ & KH & $0.3219$ & $0.0227$ & $0.0374$ & $-0.0150$ \\
  \,  & $2$ & H & $0.8695$ & $0.5106$ & $0.0569$ & $-0.4827$ \\
  \end{tabular}
  \caption{Growth rates and phase speeds for the maximum growth rate mode of the smooth profiles.``H'' stands for Holmboe. To keep parity in nomenclature, we  have  refereed even the remnants of KH (in Boussinesq free surface as well as Non-Boussinesq free surface cases) as ``KH''. These modes are strongly affected by the surface gravity waves.} 
  \label{tab:first}
  \end{center}
\end{table}

\subsection{Eigenfunction analysis of a few important modes}
\label{sec:3.5}
Eigenfunctions are useful in understanding the physical structure of the modes. The eigenfunction of the perturbation streamfunction, $\hat{\psi}$, can be obtained from the vertical velocity, $\hat{w}$ via the relation $\hat{\psi}=-\ii\hat{w}/\alpha$ ($\hat{w}$ is obtained by solving the eigenvalue problem (\ref{eq:2.6})). Since eigenfunctions are non-unique, we normalize $\hat{\psi}$ by its maximum value for reporting purposes. Apart from  $\hat{\psi}$, we have also plotted the contours of perturbation streamfunction $\tilde{\psi}=\Re \{\hat{\psi} \ee^{\ii \alpha x} \}$ for one wavelength of the disturbance.  The direction of the perturbed velocity field would be tangential to the contour lines. 
The eigenfunction of the density perturbations, $\hat{\rho}$ is also of significant interest. For plotting purposes, we have normalized it by the maximum value of the corresponding $\hat{\psi}$. The contours of perturbation density, $\tilde{\rho}$ are  plotted for one wavelength. This spatial variation reveals information about the gravity waves present in the system, their relative magnitudes and phases.
Here we have chosen to analyze two specific unstable modes, the Holmboe mode for $H/h=10$ and the SG-LV mode for $H/h=3$.

\subsubsection{ Unstable Holmboe mode for $H/h=10$}
\begin{figure}
  \centering{\includegraphics[width=5.35in]{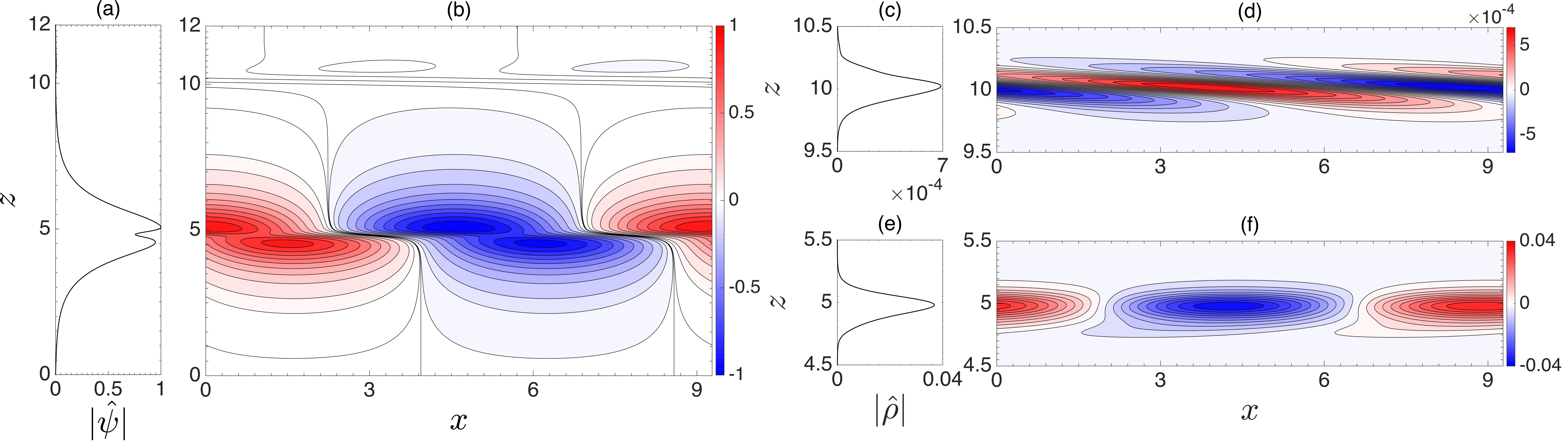}}
  \caption{Perturbation streamfunction and density for the non-Boussinesq free surface case for $H/h=10$. The plot corresponds to the maximum growth rate mode of Holmboe instability, given by $(\alpha,J)=(0.6775,0.3498)$. (a) Norm of  the perturbation streamfunction eigenfunction, $\hat{\psi}$ versus $z$, and (b) contours of the perturbation streamfunction $\tilde{\psi}$. The perturbation density characteristics near the free surface ($z=10$): (c) norm of  the perturbation density eigenfunction, $\hat{\rho}$ versus $z$ and (d) contours of the perturbation density $\tilde{\rho}$. The perturbation density characteristics near the pycnocline ($z=5$): (e) norm of  the perturbation density eigenfunction, $\hat{\rho}$ versus $z$ and (f) contours of the perturbation density $\tilde{\rho}$.}
\label{fig:3.2.5}
\end{figure}

Holmboe instability results due to the interaction between counter-propagating vorticity waves and interfacial gravity waves.  The maximum growth rate mode for  Holmboe instability corresponds to $(\alpha,J)=(0.6775,0.3498)$ in figure \ref{fig:3.2.2}(c).  The corresponding streamfunction and density perturbation contours and eigenfunctions have been plotted in figure \ref{fig:3.2.5}.   The peaks of $|\hat{\psi}|$  occur at the pycnocline (upper peak) and the lower vorticity gradient extremum (lower peak), verifying the fact that Holmboe instability arises due to the interaction between an interfacial gravity wave and a vorticity wave. We should note here that the free surface introduces a small asymmetry, favouring the ``leftward Holmboe mode'' over the ``rightward Holmboe mode''.  This is because the leftward propagating vorticity wave (which exists at the upper extrema of base vorticity gradient)  being closer to the free surface gets more affected.  Therefore, the leftward propagating Holmboe mode,  arising due to the interaction between the rightward propagating vorticity wave  and the leftward propagating interfacial gravity wave, becomes the dominant mode of the two Holmboe modes. This leftward propagating mode, with phase speed $c_r=-0.4155$ (see table \ref{tab:first}),  curves into the lower fluid region, as clearly evidenced  in figure \ref{fig:3.2.5}(f).

A few salient features of Holmboe instability in presence of a free surface are worth noticing. The value of $\hat{\rho}$ is a few orders of magnitude smaller at the free surface than that at the interface; see figures \ref{fig:3.2.5}(c)-(f). The same figures also show that the density contours at the free surface and the interface are $\pi$ phase shifted. These two features are the hallmark of the classic ``baroclinic/internal/varicose mode'' in layered flows in absence of background velocity shear; see \citet[Chapter 2]{suth2010}. 
The analogy of Holmboe instability with baroclinic mode seems appropriate because, like baroclinic mode, Holmboe instability is driven by the baroclinic torque produced at the interface. 

\subsubsection{ Unstable SG-LV mode for $H/h=3$ }
\label{sec:3.5.2}
\begin{figure}
  \centering{\includegraphics[width=5.35in]{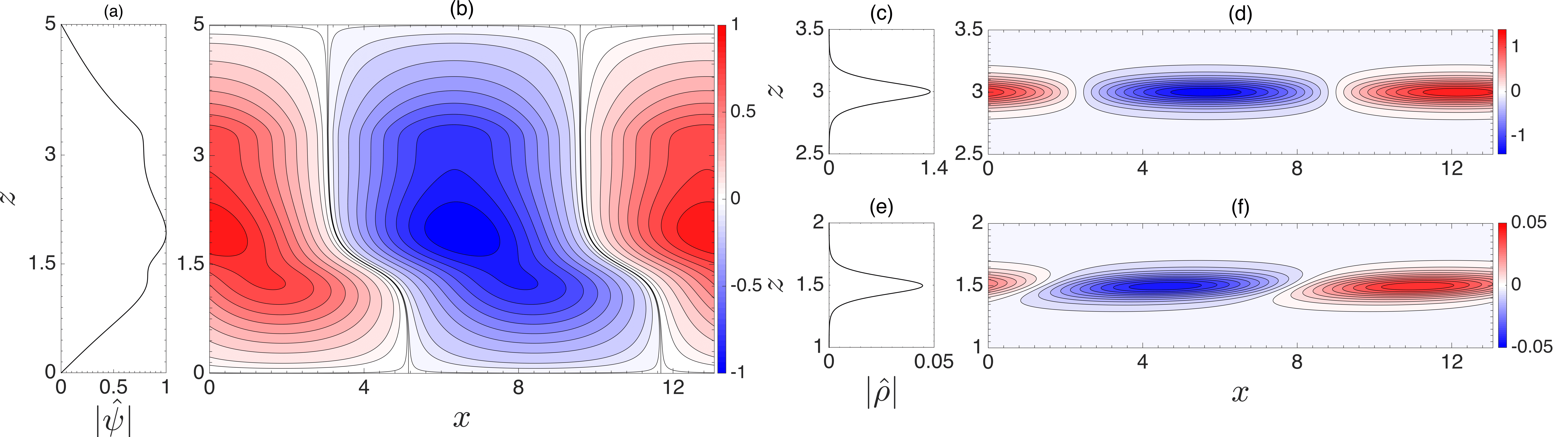}}
  \caption{ 
  As in figure \ref{fig:3.2.5} but for $H/h=3$. The plot corresponds to the maximum growth rate mode of the SG-LV instability, given by $(\alpha,J)=(0.4808,0)$.}
\label{fig:3.2.6}
\end{figure}

The maximum SG-LV mode growth rate for $H/h=3$ is given by $(\alpha,J)=(0.4808,0)$ (refer to figure \ref{fig:3.2.2}(i)).
Eigenfunctions of  perturbation streamfunction and density are plotted in figure \ref{fig:3.2.6}. 
Figures \ref{fig:3.2.6}(c)-(f) show that $|\hat{\rho}|$ has two peaks, one at the free surface and the other at the pycnocline, the former being two orders of magnitude greater than the latter. Furthermore, the surface and the interfacial gravity waves are nearly in phase. This configuration resembles the classic ``barotropic/external/sinuous mode'' in layered flows; see \citet[Chapter 2]{suth2010}.  Although from the eigenfunction plot the mechanism behind SG-LV instability is not very clear, yet a very important conclusion can be drawn - the interfacial gravity wave plays nearly no role in the instability process. In other words, the instability has to be due to the interaction between the vorticity waves present in the shear layer and the surface gravity waves.

As evident from  table \ref{tab:first}, the highest growth rate of the SG-LV mode for $H/h=3$ is comparable to KH in a nearly unbounded flow (compare with the growth rate of KH mode for $H/h=10$). The phase speed of this mode is $\approx -0.1$, and is intermediate between KH and Holmboe modes for the same $H/h$.

\section{Broken-line profiles and mechanistic picture of instability}\label{sec:4}
\subsection{Broken-line profiles and reduced order models}
\label{sec:4.1}
To complement the stability analyses of smooth profiles outlines in \S \ref{sec:3}, we undertake stability analyses of broken-line profiles in this section. \textcolor{black}{In these profiles, vorticity (vertical gradient of velocity) and density are piecewise constant while velocity is piecewise linear.}
Broken-line profiles are useful in identifying the mechanisms behind  different instabilities occurring in the flow since it allows the waves in the system to become localized.
Furthermore, using such profiles, (\ref{eq:2.3}) can be solved analytically. An equivalent form of (\ref{eq:2.3}) is given below:
\begin{equation}\label{eq:4.0}
\{\bar{\rho}[(\bar{u}-c)\hat{w}^{\prime} - \bar{u}^{\prime}\hat{w}]\}^{\prime} - \frac{\bar{\rho}^{\prime}g}{\bar{u}-c}\hat{w} - \bar{\rho}\alpha^{2}(\bar{u}-c)\hat{w} =0.
\end{equation}
One or more waves can exist at a material interface located  arbitrarily at $z=z_{i}$.
A material interface arises due to the discontinuity in the base state vorticity and/or base state density. In each layer (fluid between two consecutive interfaces), both (\ref{eq:2.3}) and (\ref{eq:4.0}) are reduced to $\bar{\rho}(\bar{u}-c)[\hat{w}^{\prime \prime}-\alpha^{2}\hat{w}]=0$. 
The discrete eigenspectrum can be obtained by assuming $(\bar{u}-c)\neq 0$, yielding
\begin{equation}\label{eq:4.0.1}
\hat{w}^{\prime \prime}-\alpha^{2}\hat{w}=0.
\end{equation}
The continuity of vertical velocity across an interface is imposed:
\begin{equation}\label{eq:4.0.2}
\llbracket \, \hat{w} \, \rrbracket=0,
\end{equation}
where $\llbracket .... \rrbracket$ denotes the difference across the interface. 
Continuity of pressure across the interface is ensured by  the dynamic condition, obtained by integrating (\ref{eq:4.0}) across the interface from $z_{i}-\Updelta z$ to $z_{i}+\Updelta z$ and letting $\Updelta z \rightarrow 0$,


\begin{equation}\label{eq:4.0.3}
\llbracket \, \bar{\rho}(\bar{u}-c)\hat{w}^{\prime}-\bar{\rho}\bar{u}^{\prime}\hat{w}-\bar{\rho}\frac{g\hat{w}}{\bar{u}-c} \, \rrbracket=0.
\end{equation}

\begin{figure}
\centering{\includegraphics[width=5.35in]{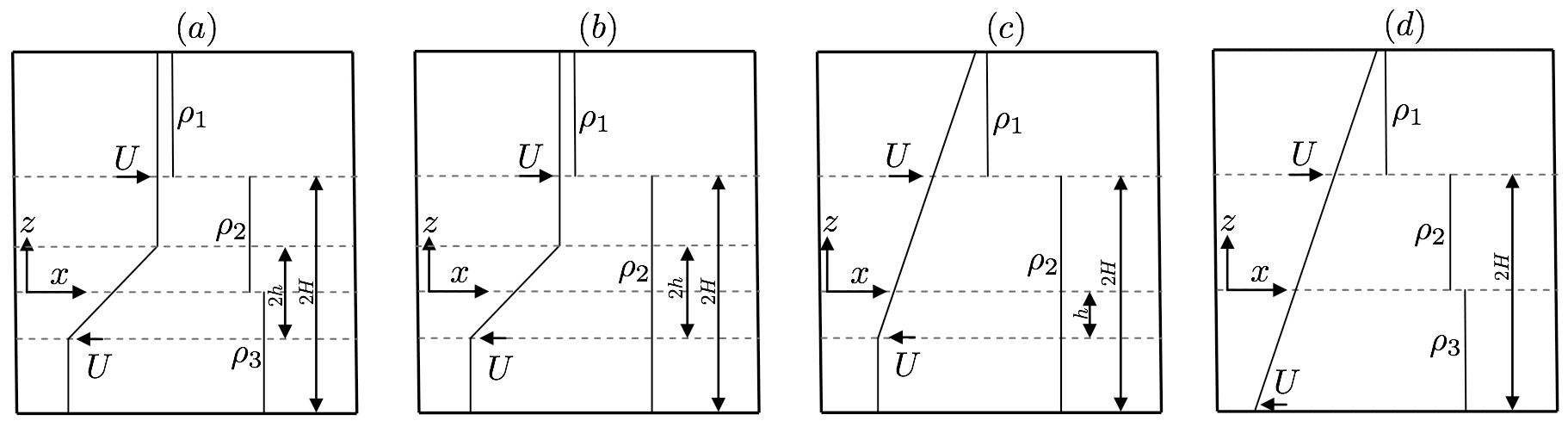}}
  \caption{Base state velocity and density plots for the broken-line profiles. (a) Boussinesq/non-Boussinesq free surface case. (b) Rayleigh/Kelvin-Helmholtz free surface model. (c) Reduced order SG-LV model. (d) Reduced order SG-IG model. }
\label{fig:4.1.1}
\end{figure}

\begin{figure}
\centering{\includegraphics[width=5.4in]{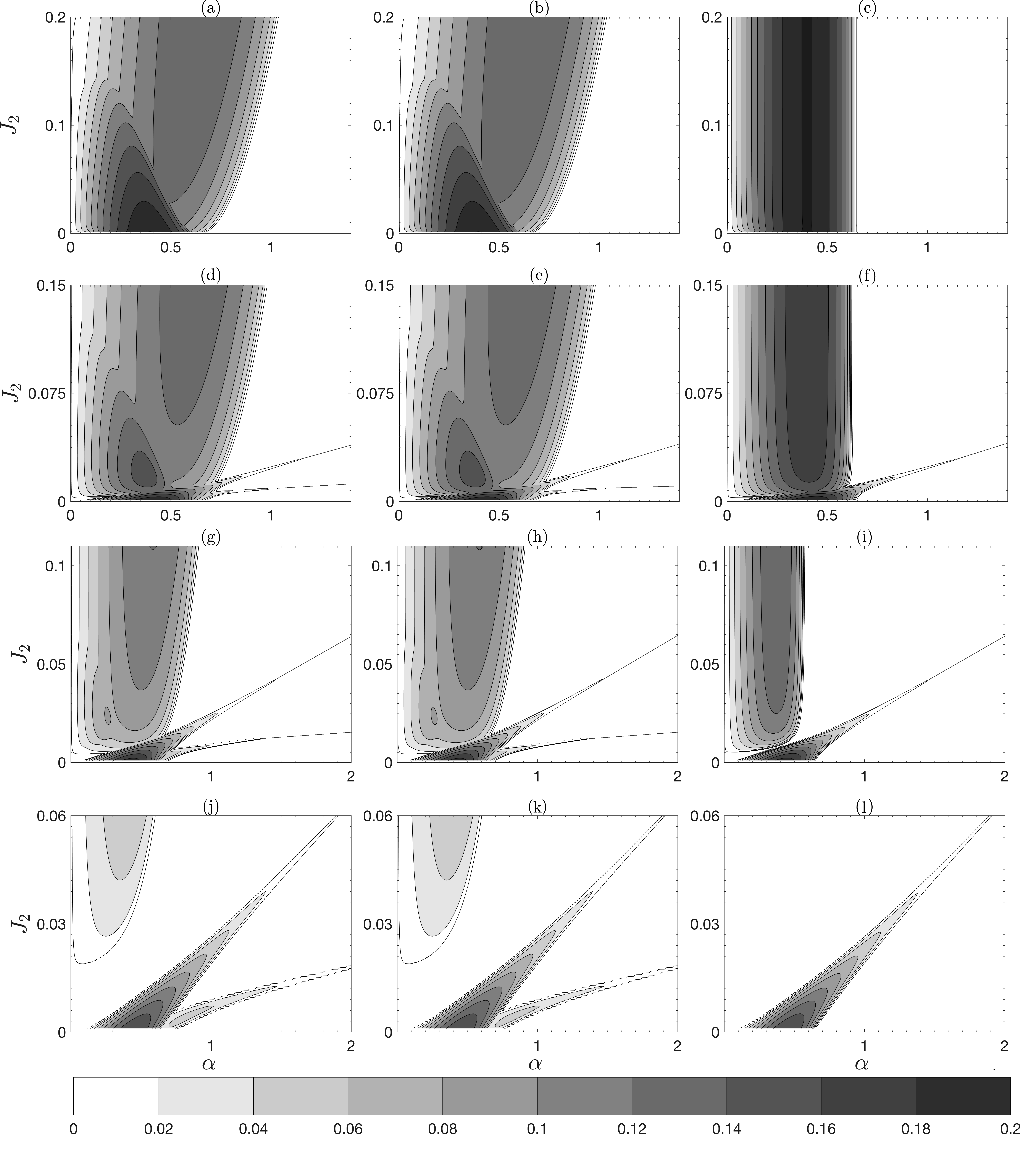}}
  \caption{Growth rate contours in the $\alpha-J_2$ plane for the broken-line profiles.  Each row represents growth rates for a given $H/h$. First row: $H/h=10$, second row: $H/h=4$, third row: $H/h=3$, and fourth row: $H/h=2$. The left, middle and right columns respectively correspond to, Boussinesq free surface, non-Boussinesq free surface and Rayleigh/Kelvin-Helmholtz free surface.}
\label{fig:4.2.2}
\end{figure}

To capture the essence of the base state profiles given in (\ref{eq:3.1.1}) and (\ref{eq:3.1.3}), the corresponding broken-line base state velocity and density profiles are given in the equations below, and are schematically shown in figure \ref{fig:4.1.1}(a):
\textcolor{black}{
\begin{subequations}
\begin{equation}\label{eq:4.1}
\bar{u}(z) = \left\{
        \begin{array}{cc}
        U & \quad z \geq h, \\ \\
             U\dfrac{ z}{h} & \quad h \geq z \geq -h, \\ \\
            -U  & \quad -h \geq z \geq -H,
        \end{array}
    \right.
\end{equation}
\begin{equation}\label{eq:4.2}
\bar{\rho}(z) = \left\{
        \begin{array}{cc}
        \rho_{1} & \quad z > H, \\ \\
            \rho_{2} & H > z > 0, \\ \\
            \rho_{3}  & 0 > z > -H.
        \end{array}
    \right.
\end{equation}
\end{subequations}
}
Base state profiles (\ref{eq:4.1})-(\ref{eq:4.2}) are first solved using the Boussinesq version of (\ref{eq:2.3}), which ignores density variation in the inertia terms. Growth rate contours corresponding to the Boussinesq free surface case are shown in the first column of figure \ref{fig:4.2.2}. From the discussion in \S \ref{sec:3.2} it was inferred that if shear is absent at the free surface (which is the case here), then the non-Boussinesq and Boussinesq modes would yield the same result. In other words, to demonstrate that the term $T_{3}$ in (\ref{eq:2.2}) is \emph{zero},  we  solve the base state profiles given in (\ref{eq:4.1})-(\ref{eq:4.2}) using  the complete equation (\ref{eq:2.3}). In this regard we solve (\ref{eq:4.0.1}) between two consecutive interfaces and use (\ref{eq:4.0.2}) and (\ref{eq:4.0.3}) as  boundary conditions. Two additional boundary conditions that are used are impervious bottom boundary and wave evanescence very far away from the free surface.
Thereby we obtain a $6$th order dispersion relation for the non-Boussinesq free surface case, which is given by $\mathcal{D}_{1}(c,\alpha;J_1;J_2;H/h)=0$. In determinant form $\mathcal{D}_{1}$ is written as

\begin{equation}\label{eq:4.3}
\mathcal{D}_{1}=
\begin{vmatrix} 
\ee^{\alpha H/h}(-P_{1}+P_{2})  & \ee^{-\alpha H/h}(P_{1}+P_{2}) & 0 & 0  & 0 & 0  \\
\ee^{\alpha}  &\ee^{-\alpha}(-2 P_{1}+1)  & 0 & 2\ee^{-\alpha}P_{1}  & 0 & 0  \\
0  & 0 & 0 & 2\alpha c  & -\dfrac{J_{2}}{c} & -2\alpha c -\dfrac{J_{2}}{c}  \\
0 & 0 & 0 & 0  & P_{3} & P_{4}   \\
\ee^{2\alpha}  & 1 & -\ee^{2\alpha} & -1  & 0 & 0  \\
0  & 0 & 1 & 1  & -1 & -1  \\
\end{vmatrix},
\end{equation}

where

\begin{multline*}
P_{1}=\alpha(1-c),\,  P_{2}=\frac{J_{1}}{1-c},\, P_{3} = -\ee^{-\alpha}\Bigg[\frac{2\alpha(1+c)\ee^{\alpha}}{\ee^{\alpha}-\ee^{\alpha\big(2H/h\,-1\big)}}+1 \Bigg],
\\ P_{4} = -\ee^{\alpha}\Bigg[\frac{2\alpha(1+c)\ee^{\alpha\big(2H/h\,-1\big)}}{\ee^{\alpha}-\ee^{\alpha\big(2H/h\,-1\big)}}+1 \Bigg]. 
\end{multline*}

Since $\rho_{1}\ll \rho_{2}$ and $\rho_{2}\approx \rho_{3}$ we have
\begin{equation*}
J_{1}=\frac{\rho_{2}-\rho_{1}}{\rho_{2}+\rho_{1}}\frac{gh}{U^{2}}\approx\frac{gh}{U^{2}},\,\,\, J_{2}=\frac{2(\rho_{3}-\rho_{2})}{\rho_{3}+\rho_{2}}\frac{gh}{U^{2}} \approx \frac{\rho_{3}-\rho_{2}}{\rho_{2}}\frac{gh}{U^{2}}. 
\end{equation*}
 The dispersion relation $\mathcal{D}_{1}=0$ is $6$th order due to the presence of six waves
 in the system, namely two vorticity waves (each present at a vorticity jump), two surface-gravity waves (present at the free surface) and two interfacial-gravity waves (present at the pycnocline). The dispersion relation is solved using the MATLAB routine `roots'. Growth rate contours for the non-Boussinesq free surface case are shown in the second column of figure \ref{fig:4.2.2}. If $T_3 = 0$ (which is our inference),  the first and second columns of figure \ref{fig:4.2.2} should be exactly the same. This can be confirmed by observing that the first and second columns of figure \ref{fig:4.2.2} are indeed identical.
 
One important objective behind performing broken-line analysis is to find qualitative as well as quantitative agreement with the corresponding smooth profile. Comparison between the third column of figure \ref{fig:3.2.2} with the second column of figure \ref{fig:4.2.2} clearly shows that the broken-line profile thoroughly captures the essence of different types of instabilities existing in the system. The quantitative agreement between smooth and broken-line profiles can be observed by comparing table \ref{tab:first} with table \ref{tab:second}. \textcolor{black}{ We also note that the thin unstable protuberances (which extends to infinity) visible in figures \ref{fig:4.2.2}(d)-(l) are not observed in figure \ref{fig:3.2.2}, since the growth rate values in these protuberances are very small (nearly zero); even a tiny amount of viscosity is enough to dampen these instabilities. In fact, even in the purely inviscid case depicted in figures \ref{fig:4.2.2}(d)-(l), these narrow unstable regions are obtained only after very careful computations. } 




The presence of six waves in the system makes it difficult to pinpoint the mechanisms leading to each type of instability. This motivated us to construct simplified or reduced order models having lesser number of waves. Since SG-LV is a new kind of instability and the mechanism leading to its formation is not yet fully understood, a reduced order set-up is constructed that has a Rayleigh/Kelvin-Helmholtz type velocity profile and a free surface. This implies that the pycnocline, and therefore the two interfacial gravity waves, have been removed from the system. This profile has been studied by both \cite{longuet1998} and \cite{bakas2009modal}, and is given in figure \ref{fig:4.1.1}(b). We have termed this model as the ``Rayleigh/Kelvin-Helmholtz free surface'' model. The base velocity and density profiles are as follows:
\textcolor{black}{ 
\begin{subequations}
\begin{equation}\label{eq:4.4}
\bar{u}(z) = \left\{
        \begin{array}{cc}
        U & \quad z \geq h, \\ \\
             U\dfrac{ z}{h} & \quad h \geq z \geq -h, \\ \\
            -U  & \quad -h \geq z \geq -H,
        \end{array}
    \right.
\end{equation}
\begin{equation}\label{eq:4.5}
\bar{\rho}(z) = \left\{
        \begin{array}{cc}
        \rho_{1} & \quad z > H, \\ \\
            \rho_{2} & H > z > -H. 
        \end{array}
    \right.
\end{equation}
\end{subequations}
}

Following a procedure similar to the non-Boussinesq free surface case a dispersion relation $\mathcal{D}_{2}(c,\alpha;J_1;H/h)=0$ can be obtained for the Rayleigh/Kelvin-Helmholtz free surface model. This dispersion relation is $4$th order (due to presence of four waves in the system), and is obtained by evaluating the following determinant and equating it to zero:

\begin{equation}\label{eq:4.3}
\mathcal{D}_{2}=
\begin{vmatrix} 
\ee^{\alpha H/h}(-P_{1}+P_{2})  & \ee^{-\alpha H/h}(P_{1}+P_{2}) & 0 & 0  & 0   \\
\ee^{\alpha}  &\ee^{-\alpha}(-2 P_{1}+1)  & 0 & 2\ee^{-\alpha}P_{1}  & 0\\
0  & 0 & 0 & -P_{5}  & P_{5}+P_{6} \\
\ee^{2\alpha} & 1 & -\ee^{2\alpha} & -1  & 0  \\
0  & 0 & -\ee^{-\alpha} & -\ee^{\alpha}  & -P_{6}  \\
\end{vmatrix}.
\end{equation}

Here $P_{5}=-2\ee^{\alpha}\alpha(1+c)$, $P_{6}=\ee^{\alpha\big(2H/h\,-1\big)}-\ee^{\alpha}$. Contours of constant growth rate for varying $H/h$ are plotted in third column of figure \ref{fig:4.2.2}. To have parity between the different growth rate contours of figure \ref{fig:4.2.2}, we have plotted the  Rayleigh/Kelvin-Helmholtz free surface model in $\alpha-J_{2}$ plane. The Rayleigh/Kelvin-Helmholtz free surface model has no $J_{2}$ inherently (since there is no pycnocline), but we can scale  $J_{1}$ by $(\rho_{3}-\rho_{2})/\rho_{2}$ to yield  $J_{2}$. For large values of $J_{2}$, the third column of figure \ref{fig:4.2.2} reveals that the Rayleigh/Kelvin-Helmholtz free surface model very closely resembles Rayleigh/KH instability.  The SG-LV branch of instability arising from the non-Boussinesq free surface case as well as the Rayleigh/Kelvin-Helmholtz free surface model are qualitatively and quantitatively similar. This implies that the interfacial gravity waves do not play any role in the SG-LV mode. We obtained the same conclusion from the eigenfunction analysis of smooth profiles in \S \ref{sec:3.5.2}. 

We emphasize here that SG-LV mode has also been observed by \cite{longuet1998} and \cite{bakas2009modal}, and is referred to as the ``branch II instability''. From the previous literature it is not clear exactly what leads to the formation of this mode, i.e.\ which waves play the central role in this instability. 
Based on the intuitive understanding of resonant interactions between counter-propagating waves, we argue that \emph{only two} waves are essential for the SG-LV instability, the leftward moving surface gravity wave (wave-$2$ in figure \ref{fig:schematic}) and the rightward moving vorticity wave existing at the lower vorticity interface \textcolor{black}{(wave-$6$ in figure \ref{fig:schematic})}. 

To prove the above-mentioned hypothesis as well as to gain a deeper understanding behind the formation of SG-LV mode, we have constructed a further reduced order model that  consists only of two oppositely propagating surface gravity waves and a rightward propagating vorticity wave. We call this the SG-LV model; see figure \ref{fig:4.1.1}(c). Base state profile for this model is given below:
\textcolor{black}{
\begin{subequations}
\begin{equation}\label{eq:4.7}
\bar{u}(z) = \left\{
        \begin{array}{cc}
        U\dfrac{2\, h}{H+h}\left(\dfrac{z}{h}\right)-U\dfrac{H-h}{H+h} & \quad z \geq -h, \\ \\
            U\dfrac{ z}{h}\,\,\,\,\,\,\,\,\,\,\,\,\,\,\,\,\,\,\,\,\,\,\,\,\,\,\,\,\,\,\,\,\,\,\,\,\,\,\,\,\,\,\,\,\,\,\,\,\,\,\,\,\,\,\,\, & \quad -h \geq z \geq -H. \\
        \end{array}
    \right.
\end{equation}
\begin{equation}\label{eq:4.8}
\bar{\rho}(z) = \left\{
        \begin{array}{cc}
        \rho_{1} & \quad z > H, \\ \\
            \rho_{2} & H > z > -H.
        \end{array}
    \right.
\end{equation}
\end{subequations}
}

\begin{figure}
\centering{\includegraphics[width=5.2in]{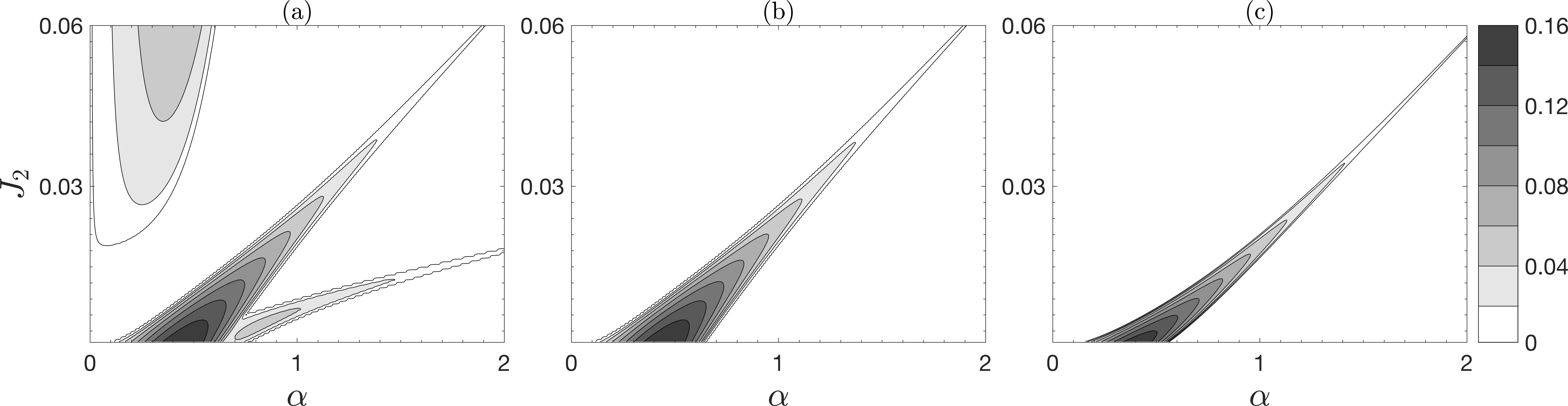}}
  \caption{Growth rate contours for $H/h=2$ case for (a) non-Boussinesq free surface, (b) Rayleigh/Kelvin-Helmholtz profile with free surface and (c) reduced order SG-LV model containing a free surface and a lower vorticity jump.  }
\label{fig:4.2.3}
\end{figure}

The dispersion relation $\mathcal{D}_{3}(c,\alpha;J_1;H/h)=0$ is obtained by equating the following determinant to zero:

\begin{equation}\label{eq:4.9}
\mathcal{D}_{3}=
\begin{vmatrix} 
\ee^{\alpha H/h}(-P_{1}+P_{2}+P_{7})  & \ee^{-\alpha H/h}(P_{1}+P_{2}+P_{7})    \\
P_{3}+\ee^{-\alpha}(1-P_{7})  &P_{4}+\ee^{\alpha}(1-P_{7})  \\
\end{vmatrix},
\end{equation}
 where $P_{7}=2h/H+1$. It is a  $3$rd order equation due to presence of three waves in the system. To have the same Doppler shift as in the Rayleigh/Kelvin-Helmholtz free surface model, we kept the velocity at the free surface to be $U$. 
A consequence of keeping the same Doppler shift is that the shear in the system changes,  hence the (non-dimensional) growth rates also change since they depend on the shear scale. The shear scale for the SG-LV model is chosen such that the maximum growth rate of this model is equal to that of the Rayleigh/Kelvin-Helmholtz free surface model. Like the Rayleigh/Kelvin-Helmholtz free surface model, $J_{1}$ has been scaled by $(\rho_{3}-\rho_{2})/\rho_{2}$ in the reduced order SG-LV model in order to get an equivalent $J_2$. This would maintain parity between different growth rate contours.  
A comparison between the non-Boussinesq free surface, Rayleigh/Kelvin-Helmholtz free surface and SG-LV models for $H/h=2$ can be made from figure \ref{fig:4.2.3}. We have chosen $H/h=2$  specifically because SG-LV mode is more prominent for small values of $H/h$.
Growth rates and stability boundaries of the SG-LV model  qualitatively as well as quantitatively agree with both the Rayleigh/Kelvin-Helmholtz free surface and the non-Boussinesq free surface cases. This clearly establishes that SG-LV results because of the interaction between the waves $2$ and $6$.

 A small but important point worth mentioning is that there is a non-zero shear at the free surface in the SG-LV model. This shear term would modify the characteristics of the surface gravity waves \citep{ehrnstrom2008linear}. This is the main reason behind the minor qualitative differences between the Rayleigh/Kelvin-Helmholtz free surface model (figure \ref{fig:4.2.3}(b)) and the SG-LV model (figure \ref{fig:4.2.3}(c)).

Apart from SG-LV mode, there is yet another instability mode that is only apparent in the broken-line profile. This is the second narrow branch observed near $\alpha=1$ and $J_{2}=0.01$ in the figure \ref{fig:4.2.2}(k). We have  termed it as the ``SG-IG mode''. To understand and isolate the essential features of this mode, we have constructed the reduced order SG-IG model, the base state profiles of which are given below (and also shown in figure \ref{fig:4.1.1}(d)):

\textcolor{black}{
\begin{subequations}
\begin{equation}\label{eq:4.10}
\bar{u}(z) = U\frac{h}{H}\bigg(\frac{z}{h}\bigg),
\end{equation}
\begin{equation}\label{eq:4.11}
\bar{\rho}(z) = \left\{
        \begin{array}{cc}
        \rho_{1} & \quad z > H, \\ \\
            \rho_{2} & H > z > 0, \\ \\
            \rho_{3}  & 0 > z > -H.
        \end{array}
    \right.
\end{equation}
\end{subequations}
}

 The dispersion relation for the SG-IG model is given by $\mathcal{D}_{4}(c,\alpha;J_1;J_2;H/h)=0$, where

\begin{equation}\label{eq:4.12}
\mathcal{D}_{4}=
\begin{vmatrix} 
\ee^{\alpha H/h}(-P_{1}+P_{2}+\dfrac{h}{H})  & \ee^{-\alpha H/h}(P_{1}+P_{2}+\dfrac{h}{H})    \\
-P_{8}-\dfrac{J_{2}}{c}  & -P_{8}\ee^{2\alpha H/h}-\dfrac{J_{2}}{c}  \\
\end{vmatrix};
\end{equation}
and $P_{8}=2\alpha c/(1-\ee^{2\alpha H/h})$.
\begin{figure}
\centering{\includegraphics[width=5.2in]{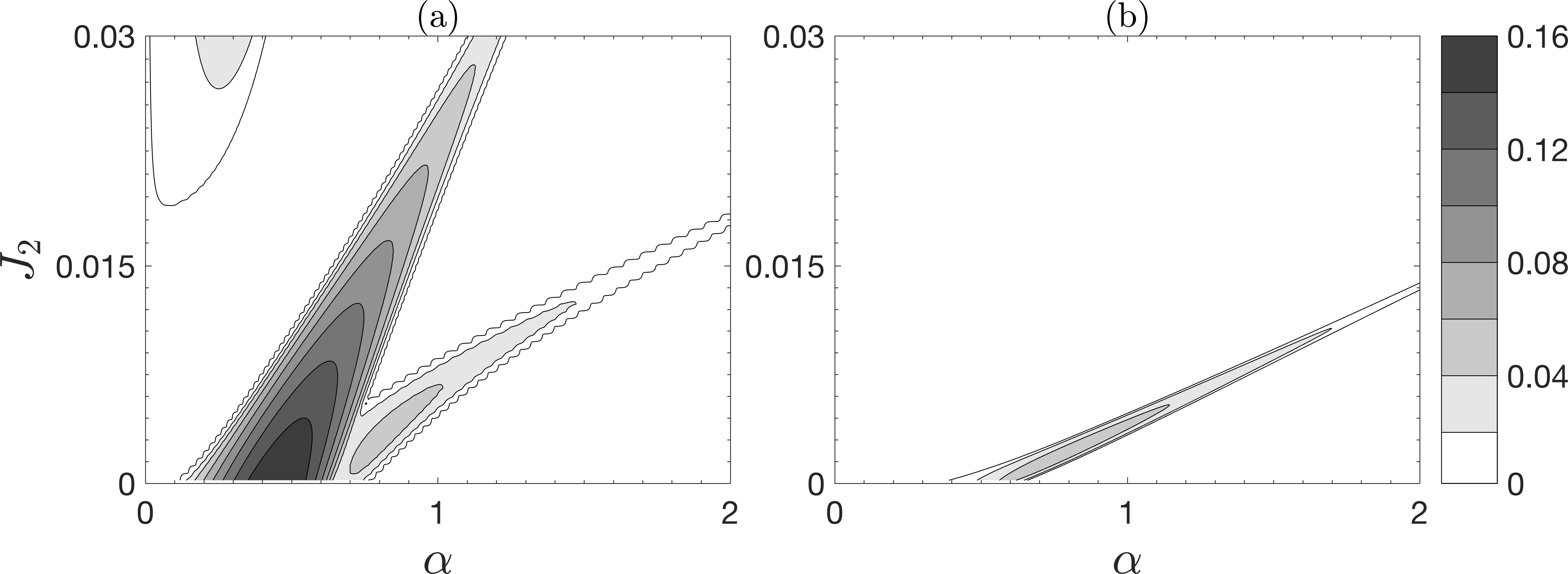}}
  \caption{Contours of constant growth rate in the $\alpha$-$J_{2}$ plane for $H/h=2$, depicting the SG-IG mode for (a) non-Boussinesq free surface and (b) reduced order SG-IG model that can only support surface gravity waves and interfacial gravity waves.  }
\label{fig:4.2.4}
\end{figure}
Four waves are present in the SG-IG model - two surface gravity waves and two interfacial gravity waves, hence $\mathcal{D}_{4}=0$ is a fourth order relation. Base state shear in the non-Boussinesq free surface case is different from the SG-IG model. Like SG-LV, here also we have chosen an appropriate shear scale for the SG-IG model so that the maximum growth rate matches that obtained for the non-Boussinesq free surface case. Figure \ref{fig:4.2.4} shows the growth rate contours for the case of non-Boussinesq free surface and the SG-IG model for $H/h=2$. The SG-IG branch of the non-Boussinesq free surface case and the reduced order SG-IG model are in qualitative and quantitative agreement. Some small differences arise because shear is present at the free surface in the reduced order SG-IG model (just like the SG-LV model).
Our analysis strongly points to the fact that the lower branch of instability appearing in the non-Boussinesq free surface case (figure \ref{fig:4.2.2}(k)) 
is due to the interaction between counter propagating surface gravity  and interfacial gravity waves (i.e.\ waves $2$ and $4$ of figure \ref{fig:schematic}).

\begin{table}
  \begin{center}
\def~{\hphantom{0}}
  \begin{tabular}{lcccccc}
      Case    & $H/h$ & Mode & $\alpha$ & $J_{2}$ & $\gamma$ & $c_{r}$    \\[3pt]
     Boussinesq/Non-Boussinesq free surface & $10$ & KH & $0.3973$ & $0$ & $0.20$ & $-0.001$    \\
       \,  & $10$ & H & $0.7975$ & $0.3973$ & $0.1422$ & $\pm 0.4667$ \\
       \,  & $4$ & SG-LV & $0.4163$ & $0$ & $0.1954$ & $-0.0942$ \\
       \,  & $4$ & SG-IG & $0.7004$ & $0.0045$ & $0.0435$ & $0.1106$ \\
        \,  & $4$ & KH & $0.3713$ & $0.0185$ & $0.1480$ & $0$ \\
         \,  & $4$ & H & $0.8235$ & $0.4388$ & $0.1415$ & $\pm 0.4844$ \\
          \,  & $3$ & SG-LV & $0.4313$ & $0$ & $0.1880$ & $-0.1945$ \\
           \,  & $3$ & SG-IG  & $0.7064$ & $0.004$ & $0.0507$ & $0.0730$ \\
            \,  & $3$ & KH & $0.2662$ & $0.023$ & $0.0814$ & $0.0128$ \\
            \,  & $3$ & H & $0.8565$ & $0.4947$ & $0.1389$ & $\pm 0.5091$ \\
             \,  & $2$ & SG-LV & $0.4573$ & $0$ & $0.1597$ & $-0.3980$ \\
            \,  & $2$ & SG-IG & $0.8425$ & $0.005$ & $0.0466$ & $0.0532$ \\  
            \,  & $2$ & H & $0.9655$ & $0.7616$ & $0.1251$ & $\pm 0.6003$ \\
    Rayleigh/Kelvin-Helmholtz with free surface &  $10$ & R & $0.4003$ & all & $0.2011$ & $-0.00088$ \\
   \,  & $4$ & R & $0.4043$ & $0.3228$ & $0.1793$ & $0$ \\
  \,  & $4$ & SG-LV & $0.4163$ & $0$ & $0.1961$ & $-0.0937$ \\
  \,  & $3$ & R & $0.3813$ & $0.9790$ & $0.1436$ & $0$ \\  
   \,  & $3$ & SG-LV & $0.4303$ & $0$ & $0.1886$ & $-0.1946$ \\
    \,  & $2$ & SG-LV & $0.4553$ & $0$ & $0.1600$ & $-0.40$ \\
     SG-LV model & $2$ & SG-LV & $0.4161$ & $0$ & $0.1542$ & $-0.5751$    \\
      SG-IG model & $2$ & SG-IG & $0.7716$ & $0.0026$ & $0.0505$ & $0.0394$    \\
     
  \end{tabular}
  \caption{Growth rates and phase speeds for the maximum growth rate mode of the broken-line profiles.}
  \label{tab:second}
  \end{center}
\end{table}

\subsection{Dispersion diagrams}
\begin{figure}
\centering{\includegraphics[width=5.2in]{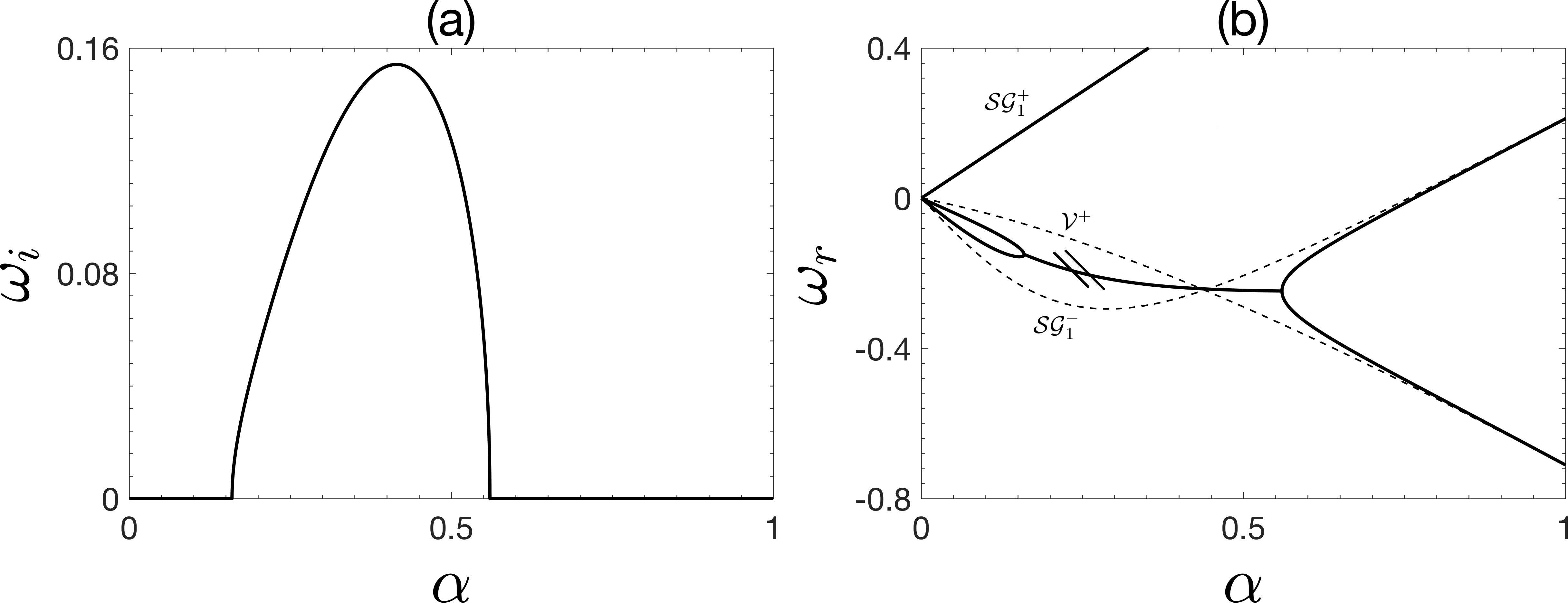}}
  \caption{Dispersion diagram for the reduced order SG-LV model for $J_2=0$ (which corresponds to the maximum growth rate case). (a) Growth rate of the unstable mode. (b) Real frequency, indicated by dark lines, are obtained by solving $\mathcal{D}_{3}=0$. The instability region is marked by two small parallel lines. The lighter lines represent the real frequencies associated with the isolated stable waves. The real frequencies of isolated waves cross near the point of maximum growth rate.}
\label{fig:4.3.1}
\end{figure}

\begin{figure}
\centering{\includegraphics[width=5.2in]{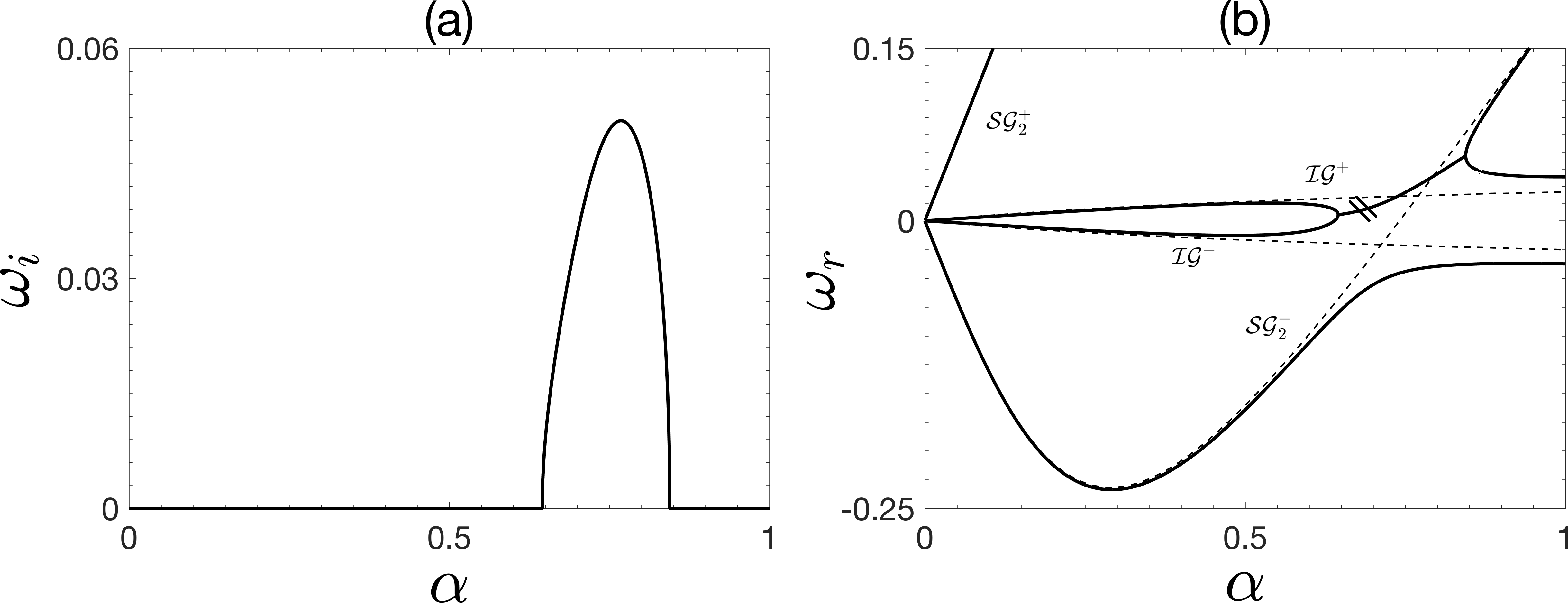}}
  \caption{ Same as figure \ref{fig:4.3.1} but   for the reduced order SG-IG model (here $J_2=0.0026$).}
\label{fig:4.3.2}
\end{figure}

A standard approach to understanding shear instabilities is via dispersion diagrams \citep{craik1988wave}, which is followed here to focus on the two new types of instability, viz.\ SG-LV and SG-IG. The dispersion diagrams of SG-LV and SG-IG models are respectively plotted in figures \ref{fig:4.3.1} and \ref{fig:4.3.2}. The dark lines correspond to the solutions of each  dispersion relation ($\mathcal{D}_{3}=0$ for SG-LV and $\mathcal{D}_{4}=0$ for SG-IG); the imaginary part of the frequency ($\omega_i$, which signifies the growth rate) is shown in figures \ref{fig:4.3.1}(a) and \ref{fig:4.3.2}(a), while the real part ($\omega_r$) is shown in figures \ref{fig:4.3.1}(b) and \ref{fig:4.3.2}(b). For a given $\alpha$, there should be $n$ roots (signifying $n$ waves) as demanded by the dispersion relation. For the SG-LV model $n=3$ while for SG-IG  $n=4$. Figure \ref{fig:4.3.1}(b) reveals that there are $3$ distinct $\omega_r$ values corresponding to each $\alpha$, except for the range $0.16<\alpha<0.56$. In this range, two roots coalesce into one. In fact, as $\alpha$  increases from $0$, two constituent waves come close together and coalesce at $\alpha=0.16$, and then bifurcate into two waves at  $\alpha=0.56$. A pair of complex conjugate roots (growing and decaying normal modes) then become a possible solution; and its presence is verified from figure \ref{fig:4.3.1}(a).
The dispersion diagram obtained by solving the Rayleigh/Kelvin-Helmholtz free surface model would produce similar dispersion diagram (number of curves would be different because SG-LV has lesser roots/waves), as can be found in \cite{longuet1998} and \cite{bakas2009modal}. 

The entire procedure applied to the SG-LV model can  also be  applied to understand the dispersion diagrams of the SG-IG model given in figures \ref{fig:4.3.2}(a)-(b). 
Furthermore, the understanding obtained from the dispersion diagrams (both SG-LV and SG-IG) can be augmented by adding the dispersion relations of individual waves in isolation. The location where two isolated waves cross each other in the $\alpha-\omega_r$ plane signifies the resonant condition (i.e.\ waves have the same $\omega_r$, which is the Doppler-shifted frequency). The dispersion relations of the isolated waves are plotted as lighter lines in figures \ref{fig:4.3.1}(b) and \ref{fig:4.3.2}(b).
Focusing on SG-LV, the dispersion relation of the isolated lower vorticity wave after appropriate Doppler shift is given by \citep{suth2010}
\begin{equation}\label{eq:SG-LV_vort}
\mathpzc{V}^{+}:\omega =-\alpha + \frac{2/3}{1+\coth(\alpha)}.
\end{equation}
%
%
Similarly, the dispersion relation for the isolated surface gravity waves affected by linear shear after appropriate Doppler shift is given by \citep{ehrnstrom2008linear}
\begin{equation}\label{eq:SG-LV_SG}
\mathpzc{SG_{1}}^{\pm}:\omega =\alpha-\frac{1}{3}\tanh(4\alpha) \pm \sqrt{\Big(\frac{1}{3}\tanh(4\alpha)\Big)^{2} + J_{1}\alpha\tanh(4\alpha)}. 
\end{equation}
 The two dispersion curves, $\mathpzc{V}^{+}$ (rightward vorticity wave) and $\mathpzc{SG_{1}}^{-}$ (leftward surface gravity wave) cross near $\alpha=0.41$ in figure \ref{fig:4.3.1}(b). This wavenumber corresponds to the most unstable mode, as can be clearly seen from  figure \ref{fig:4.3.1}(a). For large values of $\alpha$, the isolated dispersion curves asymptote to the dispersion curves obtained from solving $\mathcal{D}_{3}=0$. We note here that solutions of $\mathcal{D}_{3}=0$ give $\omega_r$ of interacting waves. Since eigenfunction of a wave decays exponentially, interaction is nearly zero when $\alpha$ is large. Hence for large $\alpha$, each curve in the dispersion diagram  basically represents an isolated wave.
The fast neutral mode obtained from the dispersion relation matches very closely with the isolated rightward surface gravity wave, $\mathpzc{SG_{1}}^{+}$, which does not interact with any other wave in the system.
\textcolor{black}{This is because the intrinsic phase speed of the rightward surface gravity wave is not opposed to the local mean flow. Hence it can not form a counter-propagating pair with any of the other waves present in the system and thus does not take part in any of the unstable modes.} 

Similar to the SG-LV model, the instability arising in the SG-IG model for  $ 0.64<\alpha<0.84$ can also be understood in terms of interacting waves in isolation. To conclusively demonstrate that the SG-IG mode arises due to the interaction between a surface gravity wave and an interfacial gravity wave, we consider them in isolation and plot their dispersion relations with lighter lines in figure \ref{fig:4.3.2}(b).
The dispersion relation for the stable isolated interfacial gravity waves is given below \citep{suth2010}: 
\begin{equation}\label{eq:SG-IG_IG}
\mathpzc{IG}^{\pm}: \omega = \pm \sqrt{\frac{ J_{2}\alpha }{2\big(1+\coth(2\alpha)\big)}}.
\end{equation}
For  stable isolated surface gravity waves in presence of shear, the dispersion relation is given by 
\begin{equation}\label{eq:SG-IG_SG}
\mathpzc{SG_{2}}^{\pm}: \omega =\alpha -\frac{1}{4}\tanh(4\alpha) \pm \sqrt{\Big(\frac{1}{4}\tanh(4\alpha)\Big)^{2}+J_{1}\alpha\tanh(4\alpha)}.
\end{equation}
The isolated waves $\mathpzc{SG_{2}}^{-}$ (leftward surface gravity wave) and $\mathpzc{IG}^{+}$ (rightward interfacial gravity wave) intersect at $\alpha\approx 0.77$ in figure \ref{fig:4.3.2}(b). This wavenumber corresponds to the most unstable mode in  figure \ref{fig:4.3.2}(a). The fast neutral mode obtained from the dispersion relation matches very closely with the isolated rightward surface gravity wave, $\mathpzc{SG_{2}}^{+}$, which does not interact with any other wave in the system. 

From figure \ref{fig:4.3.2}(b) it can be seen that the dispersion relations of isolated leftward surface gravity wave and isolated leftward interfacial gravity wave also intersect. This intersection does not lead to an exponential instability because the intrinsic phase speeds of  the leftward surface gravity wave and the leftward interfacial gravity wave are not opposite to each other. Wave interaction theory requires that the intrinsic phase speeds of exponentially growing waves have to be in opposing direction \citep{heif2005,carp2012,guha2014wave}.

Finally, we emphasize here that the identification of isolated waves which resonate to produce the observed instabilities is non-trivial. For Rayleigh, Kelvin-Helmholtz, Holmboe or Taylor-Caulfield instabilities, identification of the isolated waves are comparatively easier since in those systems, deep water internal gravity waves and/or deep water vorticity waves are present. Dispersion relations of these waves are well established. In our case, the component waves of both SG-LV and SG-IG are not that simple, and their dispersion relations are not very well known. For example, in the formation of the SG-IG mode, the ``SG wave'' is \emph{not} a deep water surface gravity wave. It turns out to be an intermediate surface gravity wave modified by linear shear, whose dispersion relation is given by (\ref{eq:SG-IG_SG}). The same is true for all the isolated waves yielding SG-LV and SG-IG modes; see (\ref{eq:SG-LV_vort})-(\ref{eq:SG-IG_SG}). Without the construction of minimal models like SG-LV (see figure \ref{fig:4.1.1}(c))   and SG-IG (see figure \ref{fig:4.1.1}(d)), identification of the key waves leading to these instabilities would not have been possible.


\section{Discussions and Conclusions}
\label{sec:5}
In this paper we have considered the effects of the vertical extent of the domain  and the air-water free surface on the stratified shear instabilities of the fluid below. In the existing literature it is often customary to replace the air-water interface by a rigid lid, thereby simplifying the problem under the  Boussinesq approximation.  However the air-water interface is a  free surface - a dynamically evolving boundary that supports surface gravity waves. Since shear instabilities  arise from resonant wave interactions, it is possible for the surface gravity waves  to resonantly interact with the other waves present in the stratified shear layer (e.g.\ vorticity waves and interfacial gravity waves). This may modify the existing instabilities, and furthermore, can lead to newer instabilities. The coupling can be made more feasible by bringing the free surface closer to the shear layer.
By considering the vertical domain to extend only up to the free surface, and furthermore, by approximating the free surface as a ``Boussinesq'' interface, \cite{longuet1998} and \cite{bakas2009modal} used broken-line profiles to understand how surface waves interact with the Rayleigh instability in the shear layer (they considered the fluid underneath to be homogeneous, i.e.\ an unstratified water body). In \S \ref{sec:3.2} we have shown that the air-water free surface can  be treated as a Boussinesq interface (by neglecting the non-Boussinesq baroclinic torque $T_3$) \emph{only} when the background shear is absent at the free surface. \textcolor{black}{ 
Since the major objective of this paper is to understand how the free surface impacts the submerged stratified shear instabilities, accurate modeling of the free surface is essential. For this reason, our vertical domain \emph{does not} end at the air-water interface (free surface) but extends up into the air region.  Including the air region may apparently seem redundant but a critical analysis reveals that when shear is present at the free surface, the boundary condition no more remains the well known dynamic boundary condition (i.e.\ the unsteady Bernoulli's equation). Thus we treat the free surface as an internal interface between air and water, and implement a free-slip, no penetration boundary up in the air region.} 

In order to capture the effect of the free surface on the stratified shear instabilities occurring at the pycnocline, we have developed a code that numerically solves the non-Boussinesq Taylor-Goldstein equation. Numerical linear stability analysis is performed on smooth base state velocity and density profiles, respectively given in (\ref{eq:3.1.1}) and (\ref{eq:3.1.3}).
First we have studied the simpler case when the air-water interface is replaced by a rigid lid, but the non-dimensional vertical height of the domain, $H/h$, is  varied. Our results are in agreement with similar previous studies by \cite{hazel1972numerical} and \cite{haigh1999symmetric}. The maximum growth rate of the KH instability decreases significantly from $0.19$ to $0.096$ on reducing $H/h$ from $10$ to $2$. However, as predicted by \cite{haigh1999symmetric},  Holmboe instability is little affected by changes in $H/h$. If rigid-lid approximation is not implemented, i.e.\ the air-water interface is treated as a non-Boussinesq free surface, the stability boundaries are found to change drastically on decreasing $H/h$. Only for large values of $H/h$, rigid lid becomes a valid approximation since, in this scenario, the free surface has practically no effect on the shear layer. Hence we infer that  rigid-lid approximation is somewhat misleading and therefore should be applied with caution. 

For higher values of $H/h$, the most unstable mode (for all bulk Richardson numbers) is due to the Rayleigh/KH instability, which arises from the coupling between two vorticity waves, each existing at an extrema of the base vorticity gradient (for broken-line profiles, this would translate to the jump in the base vorticity profile). However, as $H/h$ is decreased, a new mode of instability appears. This mode, which we refer to as the SG-LV mode, becomes the most unstable one, even surpassing KH. In fact, for very low values of  $H/h$, KH is nearly non-existent, and the dominant instability is due to SG-LV; see table \ref{tab:first}. We note here that similar instability has been observed by \cite{longuet1998} and \cite{bakas2009modal} while using broken-line profiles. Holmboe instability is found to be quite resilient to variations in $H/h$, similar to what observed in the rigid-lid case. 
Hence it can be concluded that rigid lid is a valid approximation even for shallow domains if one is \emph{only} interested in studying Holmboe instability. 

The eigenfunction analysis performed in \S \ref{sec:3.5} reveals that Holmboe instability is analogous to the baroclinic mode in two-layered flows - the surface and interface are $\pi$ shifted in phase, and furthermore, the surface elevation is insignificant in comparison to that of the interface. This is probably the reason why Holmboe instability is relatively unaffected by variations in $H/h$. 
The SG-LV mode, on the other hand, is analogous to the barotropic mode in two-layered flows - the surface and interface (pycnocline) are nearly in phase, and the interface elevation is insignificant in comparison to that of the surface. This also implies that the pycnocline plays an insignificant role in this instability. 

To complement the numerical stability analysis of the smooth profiles, and furthermore, to obtain a simplified understanding of the instabilities in the system, we have also performed  stability analyses of  broken-line profiles. A qualitative as well as quantitative agreement between the stability analysis of smooth and broken-line profiles is observed; compare figure \ref{fig:3.2.2} with figure \ref{fig:4.2.2} and table \ref{tab:first} with table \ref{tab:second}. The SG-LV mode is clearly observed as a distinct branch of instability which arises as $H/h$ is decreased (i.e.\ the depth is made shallower). 
An additional type of instability is observed in the broken-line profile that is hidden in the analysis of smooth profile (we refer to it as the SG-IG mode). Like SG-LV, the SG-IG instability is also observed when the depth is quite shallow (e.g.\ $H/h \lesssim 4$).
The broken-line profile paved the idea to construct a few reduced order broken-line models so as to underpin the wave interactions that lead to different instabilities. We have constructed three reduced order models, viz.\ Rayleigh/Kelvin-Helmholtz with free surface, SG-LV and SG-IG; see figure \ref{fig:4.1.1}.

Through the reduced order SG-LV model we have conclusively shown that the SG-LV mode arises primarily because of the interaction between the leftward propagating surface gravity wave and the rightward propagating vorticity wave, i.e.\ interaction of waves $2$ and $6$ of figure \ref{fig:schematic}. The understanding is augmented by studying the dispersion diagrams of the full system, as well as the stable isolated waves that are suspected to be behind these instabilities. The crossing of the dispersion curves of the isolated stable waves in the $\alpha-\omega_r$ plane corresponds to the unstable region in the $\alpha-\omega_i$ plane, confirming our initial assumption. We emphasize here that, although this mode of instability has been observed previously by \cite{longuet1998} and \cite{bakas2009modal} (they refer to it as the branch II instability), the fundamental reason behind it was not clearly known. Neither  was it known whether such instabilities exist in smooth profiles, and if yes, then how well does it compare with the broken-line counterpart. In this paper we have been able to address all these points conclusively. 


An approach similar to SG-LV is taken to understand the SG-IG mode. A reduced order SG-IG model is constructed to underpin the instability mechanism. Finally, with the help of dispersion diagrams, we confirmed that the SG-IG mode is indeed an interaction of the leftward surface gravity wave (wave-$2$) and rightward interfacial gravity wave (wave-$4$). This mode, to the best of our knowledge, \textcolor{black}{has not been reported previously in the literature}.


In summary, we have performed a comprehensive study on the effect of free surface on the stratified shear instabilities underneath. For shallow flows, i.e., when the free surface is relatively closer to the shear layer, the free surface significantly affects the ensuing shear instabilities. The surface gravity waves resonate with the different waves present at the shear layer, thereby modifying the ``well-known'' instabilities, and more importantly, giving rise to two new instabilities. These important dynamics won't be captured if the air-water interface is modeled as a rigid lid.  When  shear is present at the free surface, the non-Boussinesq baroclinic torque may become significant and therefore can strongly affect the stability characteristics. Moreover, the non-Boussinesq baroclinic torque is absent when  shear is absent at the free surface. Thus, although there is an $\mathcal{O}(1)$ density variation at the free surface, a Boussinesq like approximation is sufficient (that is, one can only consider the gravitational part of the baroclinic torque) when free surface has no background shear.  Finally, we point out here that for analytical simplicity, we have considered the pycnocline to be at the mid-depth of the water body. However in real aquatic environments, the pycnocline is usually closer to the free surface. Thus the surface waves can have a more stronger influence on the submerged shear instabilities than that considered in this paper.  Future experimental studies and/or  Direct numerical simulation (DNS) may be able to shed more light into SG-LV and SG-IG instabilities, especially their non-linear evolution and three dimensional structures.

\appendix
\section{Derivation of Non-Boussinesq Taylor-Goldstein Equation }\label{appA}
A full derivation of the non-Boussinesq viscous diffusive Taylor-Goldstein equation is given here.
A 2D flow in the $x$-$z$ plane is considered. The horizontal, $x$ component of velocity is given by $u$  while the vertical, $z$ component of velocity is given by $w$. The dynamic viscosity (which is assumed constant) and mass density of fluid are given by $\mu$ and $\rho$ respectively. Acceleration due to gravity is given by $g$, while  $t$ denotes time. The governing equations of the problem are given below.\\
Incompressible continuity equation:
\begin{equation}\label{eq:A1}
\frac{\partial u}{\partial x}+\frac{\partial w}{\partial z}=0.
\end{equation}
Navier-Stokes equation for the $x$-momentum:
\begin{equation}\label{eq:A2}
\rho \Big(\frac{\partial u}{\partial t}+u\frac{\partial u}{\partial x}+w\frac{\partial u}{\partial z}\Big)=-\frac{\partial p}{\partial x}+\mu\Big(\frac{\partial^{2}u}{\partial x^{2}}+\frac{\partial^{2}u}{\partial z^{2}}\Big).
\end{equation}
Navier-Stokes equation for the $z$-momentum: 
\begin{equation}\label{eq:A3}
\rho \Big(\frac{\partial w}{\partial t}+u\frac{\partial w}{\partial x}+w\frac{\partial w}{\partial z}\Big)=-\frac{\partial p}{\partial z}-\rho g+ \mu\Big(\frac{\partial^{2}w}{\partial x^{2}}+\frac{\partial^{2}w}{\partial z^{2}}\Big).
\end{equation}
Advection diffusion of the stratifying agent:
\begin{equation}\label{eq:A4}
\frac{\partial \theta}{\partial t}+u\frac{\partial \theta}{\partial x}+w\frac{\partial \theta}{\partial z}=\eta\Big(\frac{\partial^{2}\theta}{\partial x^{2}}+\frac{\partial^{2}\theta}{\partial z^{2}}\Big).
\end{equation}
Equation of state relating stratifying agent to density:
\begin{equation}\label{eq:A5}
\rho =\rho_{0}[1-\beta(\theta-\theta_{0})]. 
\end{equation}
In equations (\ref{eq:A4}) and (\ref{eq:A5}), $\theta$ is the stratifying agent like temperature or salinity, $\eta$ is the  (constant)  molecular diffusivity of the stratifying agent, while $\beta$ is the linear coefficient relating density to changes in stratifying agent.  The quantities $\rho_{0}$ and $\theta_{0}$ are the base state density and some reference value of the stratifying agent concentration. Combining equations (\ref{eq:A4}) and (\ref{eq:A5}) we obtain an advection diffusion equation for mass density given by
\begin{equation}\label{eq:A6}
\frac{\partial \rho}{\partial t}+u\frac{\partial \rho}{\partial x}+w\frac{\partial \rho}{\partial z}=\kappa\Big(\frac{\partial^{2}\rho}{\partial x^{2}}+\frac{\partial^{2}\rho}{\partial z^{2}}\Big).
\end{equation}
We assume a base state that depends only on $z$, and is given by $u=\bar{u}(z)$, $w=0$, $p=\bar{p}(z)$ and $\rho=\bar{\rho}(z)$. The base state is also assumed to be under hydrostatic balance: $d\bar{p}/dz=-\bar{\rho}g$.  Infinitesimal perturbations,   denoted by $\tilde{f}$  (where $f$ is a placeholder variable), are added to the base state and then substituted in  (\ref{eq:A1})-(\ref{eq:A3}), and (\ref{eq:A6}). These equations after linearization yields

\begin{equation}\label{eq:A11}
\frac{\partial\tilde{u}}{\partial x}+\frac{\partial\tilde{w}}{\partial z}=0,
\end{equation}
\begin{equation}\label{eq:A12}
\bar{\rho} \left(\frac{\partial \tilde{u}}{\partial t}+\bar{u}\frac{\partial \tilde{u}}{\partial x}+\tilde{w}\frac{d\bar{u}}{dz}\right)=-\frac{\partial \tilde{p}}{\partial x} +\mu\Big(\frac{\partial^{2}\tilde{u}}{\partial x^{2}}+\frac{\partial^{2}\tilde{u}}{\partial z^{2}}\Big),
\end{equation}
\begin{equation}\label{eq:A13}
\bar{\rho} \left(\frac{\partial \tilde{w}}{\partial t}+\bar{u}\frac{\partial \tilde{w}}{\partial x}\right)=-\frac{\partial \tilde{p}}{\partial z}-\tilde{\rho}g+\mu\Big(\frac{\partial^{2}\tilde{w}}{\partial x^{2}}+\frac{\partial^{2}\tilde{w}}{\partial z^{2}}\Big),
\end{equation}
\begin{equation}\label{eq:A14}
\frac{\partial \tilde{\rho}}{\partial t}+\bar{u}\frac{\partial \tilde{\rho}}{\partial x}+\tilde{w}\frac{d\bar{\rho}}{dz}=\kappa\Big(\frac{\partial^{2}\tilde{\rho}}{\partial x^{2}}+\frac{\partial^{2}\tilde{\rho}}{\partial z^{2}}\Big).
\end{equation}
We have assumed  perturbations of the normal mode form $\tilde{f}(x,z,t)=\hat{f}(z)\ee^{\ii \alpha(x-ct)}$, where $\alpha$ is the wavenumber and  $c=c_r+\ii c_i$ is the complex phase speed. 
Such form on substitution in  (\ref{eq:A11})-(\ref{eq:A14}) yields
\begin{equation}\label{eq:A15}
\ii\alpha \hat{u}+\hat{w}^{\prime}=0,
\end{equation}
\begin{equation}\label{eq:A16}
\bar{\rho}\left[\ii \alpha(\bar{u}-c) \hat{u}+\hat{w}\bar{u}^{\prime}\right]=-\ii \alpha \hat{p} +\mu(\hat{u}^{\prime\prime}-\alpha^{2}\hat{u}),
\end{equation}
\begin{equation}\label{eq:A17}
\bar{\rho}\left[\ii \alpha(\bar{u}-c) \hat{w}\right]=-\hat{p}^{\prime}-\hat{\rho}g+\mu(\hat{w}^{\prime\prime}-\alpha^{2}\hat{w}),
\end{equation}
\begin{equation}\label{eq:A18}
\ii \alpha(\bar{u}-c) \hat{\rho}+\hat{w}\bar{\rho}^{\prime}=\kappa(\hat{\rho}^{\prime\prime}-\alpha^{2}\hat{\rho}).
\end{equation}
Here $^{\prime}$ denotes $d/dz$. Making the substitution $\hat{u}=\ii \hat{w}^{\prime}/\alpha$ we get
\begin{equation*}
\bar{\rho}[-(\bar{u}-c)\hat{w}^{\prime}+\bar{u}^{\prime}\hat{w}]=-\ii\alpha\hat{p}+\frac{\ii}{\alpha}\mu[\hat{w}^{\prime\prime\prime}-\alpha^{2}\hat{w}^{\prime}].
\end{equation*}
Taking the total derivative of the above  equation with respect to $z$ we obtain,
\begin{equation}\label{eq:A19}
\bar{\rho}^{\prime}[-(\bar{u}-c)\hat{w}^{\prime}+\bar{u}^{\prime}\hat{w}]+\bar{\rho}[-\bar{u}^{\prime}\hat{w}^{\prime} -(\bar{u}-c)\hat{w}^{\prime\prime} +\bar{u}^{\prime\prime}\hat{w}+\bar{u}^{\prime}\hat{w}^{\prime}]=-\ii\alpha\hat{p}^{\prime}+\frac{\ii}{\alpha}\mu[\hat{w}^{\prime\prime\prime\prime}-\alpha^{2}\hat{w}^{\prime\prime}].
\end{equation}
Expressing $\hat{p}^{\prime}$ in terms of other variables from (\ref{eq:A17}) we get
\begin{equation*}
	\hat{p}^{\prime}=-\ii\bar{\rho}\hat{w} \alpha(\bar{u}-c) -\hat{\rho}g+\mu(\hat{w}^{\prime\prime}-\alpha^{2}\hat{w}).
\end{equation*}
$\hat{p}^{\prime}$ from the previous equation can be substituted in (\ref{eq:A19}) to give
\begin{equation}\label{eq:A20}
\bar{\rho}^{\prime}[-(\bar{u}-c)\hat{w}^{\prime}+\bar{u}^{\prime}\hat{w}]+\bar{\rho}[-(\bar{u}-c)\hat{w}^{\prime\prime}+\alpha^{2}(\bar{u}-c)\hat{w}+ \bar{u}^{\prime\prime}\hat{w}]=\ii\alpha\hat{\rho}g+\frac{\ii}{\alpha}\mu[\hat{w}^{\prime\prime\prime\prime}-2\alpha^{2}\hat{w}^{\prime\prime}+\alpha^{4}\hat{w}],
\end{equation}
\begin{equation}\label{eq:A21}
\ii \alpha(\bar{u}-c) \hat{\rho}+\hat{w}\bar{\rho}^{\prime}=\kappa(\hat{\rho}^{\prime\prime}-\alpha^{2}\hat{\rho}).
\end{equation}
Equations (\ref{eq:A20})-(\ref{eq:A21}) form the \emph{non-Boussinesq viscous diffusive Taylor-Goldstein equations}. Inviscid limit gives rise to $\mu \rightarrow 0$, while the non-diffusive limit yields $\kappa \rightarrow 0$. In the limiting condition of inviscid and non-diffusive flow, we obtain
\begin{equation}\label{eq:A22}
\bar{\rho}^{\prime}[-(\bar{u}-c)\hat{w}^{\prime}+\bar{u}^{\prime}\hat{w}]+\bar{\rho}[-(\bar{u}-c)\hat{w}^{\prime\prime}+\alpha^{2}(\bar{u}-c)\hat{w}+ \bar{u}^{\prime\prime}\hat{w}]=\ii\alpha\hat{\rho}g,
\end{equation}
\begin{equation}\label{eq:A23}
\ii \alpha(\bar{u}-c) \hat{\rho}+\hat{w}\bar{\rho}^{\prime}=0.
\end{equation}
Combining (\ref{eq:A22}) and (\ref{eq:A23}) we get
\begin{equation}\label{eq:A24}
\bar{\rho}^{\prime}\Big[(\bar{u}-c)\hat{w}^{\prime}- \bar{u}^{\prime}\hat{w} - \frac{g}{\bar{u}-c}\hat{w} \Big]  +  \bar{\rho}\Big[(\bar{u}-c)(\hat{w}^{\prime\prime}-\alpha^{2}\hat{w}) - \bar{u}^{\prime\prime}\hat{w} \Big]=0.
\end{equation}
The above equation is the \emph{non-Boussinesq Taylor-Goldstein equation} (inviscid and non-diffusive limit of (\ref{eq:A20})-(\ref{eq:A21})), and \textcolor{black}{is same as the one obtained by \cite{barros2011holmboe}, \cite{barros2014elementary} and \cite{carpenter2017}.}

\bibliographystyle{jfm}
\bibliography{references}

\end{document}